\documentclass[aps,prb,twocolumn,groupedaddress,showpacs]{revtex4}
\usepackage{amssymb}
\usepackage{amsmath}
\usepackage{amsthm}
\usepackage{epsfig}
\usepackage{bm}
\usepackage{mathrsfs}
\usepackage{cleveref}
\usepackage{graphicx}
\usepackage{yfonts}
\usepackage[caption=false]{subfig}
\usepackage{lipsum}
\allowdisplaybreaks[4]
\crefname{figure}{Fig.}{Figs.}
\crefname{equation}{Eq.}{Eqs.}
\usepackage{xcolor}

\begin{document}

\title{Universal thermodynamics of the one-dimensional attractive Hubbard model}
\author{Song Cheng}
\affiliation{State Key Laboratory of Magnetic Resonance and Atomic and Molecular Physics,
Wuhan Institute of Physics and Mathematics, Chinese Academy of Sciences, Wuhan 430071, China}
\affiliation{University of Chinese Academy of Sciences, Beijing 100049, China.}
\affiliation{Department of Theoretical Physics, Research School of Physics and Engineering,
Australian National University, Canberra ACT 0200, Australia}

\author{Yi-Cong Yu}
\affiliation{State Key Laboratory of Magnetic Resonance and Atomic and Molecular Physics,
Wuhan Institute of Physics and Mathematics, Chinese Academy of Sciences, Wuhan 430071, China}
\affiliation{University of Chinese Academy of Sciences, Beijing 100049, China.}

\author{M. T. Batchelor}
\affiliation{Centre for Modern Physics, Chongqing University, Chongqing 400044, China}
\affiliation{Department of Theoretical Physics,
Research School of Physics and Engineering,
Australian National University, Canberra ACT 0200, Australia}
\affiliation{Mathematical Sciences Institute,
Australian National University, Canberra ACT 0200, Australia}

\author{Xi-Wen Guan}
\email[]{xiwen.guan@anu.edu.au}
\affiliation{State Key Laboratory of Magnetic Resonance and Atomic and Molecular Physics,
Wuhan Institute of Physics and Mathematics, Chinese Academy of Sciences, Wuhan 430071, China}
\affiliation{Department of Theoretical Physics,
Research School of Physics and Engineering,
Australian National University, Canberra ACT 0200, Australia}

\affiliation{Center for Cold Atom Physics, Chinese Academy of Sciences, Wuhan 430071, China}

\date{\today}


\begin{abstract}
The one-dimensional (1D) Hubbard model, describing electrons on a lattice with an on-site repulsive interaction,
provides a paradigm for the physics of quantum many-body phenomena.
%
%However, the Hubbard model with an attractive interaction is notoriously difficult to study due to the
%presence of charge bound states of multi-electrons on the 1D  lattice.
%
Here by solving the thermodynamic Bethe ansatz equations we study the universal thermodynamics, quantum criticality and  magnetism
of the 1D attractive Hubbard model.
We show  that the compressibility and the susceptibility  of the Fulde-Ferrell-Larkin-Ovchinnikov (FFLO)-like state obey simple additivity rules at low temperatures,
indicating an existence of two  free quantum fluids.
The magnetic properties, such as magnetization and susceptibility, reveal three physical regions: quantum fluids  at low temperatures,
a non-Fermi liquid at high temperatures and the quantum fluid to non-Fermi liquid crossover in between.
The lattice interaction is seen to significantly influence the nature of the FFLO-like state in 1D.
Furthermore, we show  that the dimensionless Wilson ratio provides an ideal parameter to map out the various phase boundaries and to
characterize the two  free  fluids   of the FLLO-like state.
The quantum scaling functions for the thermal and magnetic properties yield the same dynamic critical exponent $z=2$ and
correlation critical exponent $\nu=1/2$ in the quantum critical region whenever a phase transition occurs.
Our results provide a rigorous understanding of quantum criticality and free fluids of many-body systems on a 1D lattice.

 \end{abstract}

\maketitle

\clearpage
%\tableofcontents

\section{Introduction}

How to capture the essential features of many-body physics through a simple model is always of great importance in condensed matter physics.
In this regard,  the Hubbard model~\cite{Hubbard1963} has long provided an active area of research since it was put forward
as an instance of a Mott insulator and later considered as a potential high-$T_c$ superconductor.
 The Hubbard model has thus become a prototypical strongly correlated system which provides rich many-body phenomena,  such as a Mott transition,
 superconductivity, spin-charge separation and a Fulde-Ferrell-Larkin-Ovchinnikov (FFLO) state.
However, the Hubbard model, as a simplification of interacting fermions on realistic lattices, can be analytically resolved in neither two-dimensions (2D) nor three-dimensions (3D).
The one-dimensional (1D) case within a single band is integrable, firstly solved by Lieb and Wu in terms of  the
Yang-Baxter equation~\cite{Yang:1967,Baxter:1972} and the nested Bethe ansatz~\cite{Lieb1968} (see Ref.~\onlinecite{Ess05} for an extensive review).
More specifically, since Lieb and Wu's seminal work, the 1D repulsive Hubbard model has been investigated in various aspects,
including, but not restricted to, thermodynamic properties in the ground state~\cite{MT1969s,MT1971,Shiba1972,Krivnov1975,Usuki1989,Woyna1991},
low-lying excitations~\cite{Coll1974,Klumper1990,Ovchi1970,Woyna1982s,Woyna1983,Degu2000,Ess1994a,Ess1994b},
finite temperature thermodynamics~\cite{MT1972,MT1974,Lee1988,Okiji1989,Sacra1995,Degu2000}
and correlation functions~\cite{Frahm1990,Woyna1987,Ogata1990,Parola1990,Penc1995,Gohmann1998,Ess2000,Bogoliubov1988,DMRG,QMC,SC}.

The thermodynamics of the 1D Hubbard model  is accessible through two alternative approaches -- the thermodynamic Bethe ansatz (TBA) equations \cite{MT1972}
and the quantum transfer matrix method \cite{Klumper1996}.
The former is established on the so-called `string  hypothesis' and Yang-Yang grand canonical ensemble approach \cite{Yang1969},
whereas  the latter  stems from the lattice path integral formulations for the partition function \cite{Koma1987}.
 In principle, the low-lying excitations can be constructed with the help of the TBA equations in the zero temperature limit and by the logarithm
 of the Lieb-Wu equations \cite{Degu2000,Ess1994a,Ess1994b}.
Despite these systematic approaches and other methods  employed for the study of the
ground state properties \cite{Lieb1968,MT1969s,MT1971,Shiba1972,Krivnov1975,Usuki1989,Woyna1991}
and low-lying excitations \cite{Coll1974,Klumper1990,Ovchi1970,Woyna1982s,Woyna1983},
a complete understanding of the universal thermodynamics and quantum criticality of the 1D Hubbard model has not yet been achieved.
The key reason for preventing the solution of this problem is the  difficulty of finding  a suitable generating function for the equation of state at low temperatures.

On the other hand, the correlation functions are also extremely difficult to calculate  directly using the Bethe wave function.
For  a 1D conformally invariant  system, the critical exponents determining the power law decay of correlation functions
are connected with finite-size corrections to the ground state energy \cite{Cardy1984,Aff1986,Blote1986}.
The 1D repulsive Hubbard model is conformally invariant only in the vicinity of Fermi points.
 The conformal field theory (CFT) approach provides one method to obtain the asymptotics of correlation functions \cite{Belavin1984}.
The low-lying excitations provide a practicable opportunity for  an investigation of long distance asymptotics of correlation functions \cite{Frahm1990},
where the finite-size corrections are  accessible through the Bethe ansatz method \cite{Woyna1987}.
However, difficulties involved in the actual calculations of correlation functions usually  prevent full
access to the many-body correlations \cite{Ogata1990,Parola1990,Penc1995,Gohmann1998,Ess2000}.

The mechanism of Cooper pairing in the 1D attractive Hubbard model has attracted attention \cite{Bogoliubov1988} due to the discovery of high-temperature superconductors.
 In particular, the FFLO-like pair correlation and spin correlations are consequently  investigated by various methods,
 such as density-matrix renormalization group \cite{DMRG}, quantum Monte Carlo \cite{QMC} and CFT \cite{ZhaoE:2008,SC}.
 The nature of the FFLO-like pair correlation was predicted in expansion dynamics of the attractive Hubbard model  trapped in 1D \cite{Kajala:2011}.
Very recently,  trapping cold atoms on  optical lattices becomes a promising  method to simulate the many-body physics of  the Hubbard model \cite{Singha2011,Hart2015,Greif2016,Parsons2016,Lawrence2016,Boll2016,Zhang:2015,Mazurenko:2017}.
In particular,  ultracold  atoms offer an ideal platform for testing results predicted from  1D exactly solvable  models \cite{Boll2016}.

It is understood that the macroscopic behaviour of 1D materials, such as  the spin compound Cu(C${}_4$H${}_4$N${}_2$)(NO${}_3$)${}_2$ \cite{Kono:2015}
and the heavy fermion material YbNi${}_4$P${}_2$ \cite{Krellner:2016}, demonstrates a type of 3D Fermi liquid behaviour \cite{Carmelo:1992,Shaginyan:2016}.
The motivation of the present work is to provide understanding of free fluid  nature and quantum criticality in the context of the 1D attractive Hubbard model.
Firstly, the 1D attractive Hubbard model plays an important role in understanding many-body phenomena such as superconductivity,
BEC-BCS crossover and FFLO-like  correlation \cite{DMRG},
with several publications touching upon it \cite{MT1969s,Krivnov1975,Woyna1983,Lee1988,Bogoliubov1988,Woyna1991,Ess1994a,Ess1994b,Sacra1995}.
 Secondly, one expects to find universal behaviour for this model, including thermodynamics, quantum criticality and Luttinger liquid properties.
 Thirdly, regarding  the complicated FFLO state, it is highly desirable to obtain simple rules to describe the nature of quantum liquids  in the attractive Hubbard model.
 Last, but not least, the interplay of this work with experiments with ultracold atoms \cite{Singha2011,Hart2015,Greif2016,Parsons2016,Lawrence2016,Boll2016,Zhang:2015,Mazurenko:2017} may broaden our knowledge of many-body physics through 1D exactly solvable models.

This paper is organized as follows. In section II, we present  a derivation of the TBA equations for the 1D attractive Hubbard model and
determine the ground state phase diagram.
In section III, we derive the equation of state in the strong coupling regime.
In section IV, using  the equation of state, we obtain various analytical results for the thermodynamics and magnetism which are relevant to experimental study.
We also investigate quantum criticality and obtain the universal scaling forms of thermodynamic quantities.
In section V, we demonstrate  the free fluid nature of the FFLO phase through  the simple additivity rules of  the thermodynamic quantities.
We find that  the compressibility Wilson ratio is  very powerful in identifying the Fermi liquid/Tomonaga-Luttinger liquid phases in the low temperature phase diagram.
The last section VI is reserved for  a summary  and conclusion.

We conclude this section by noting that this article provides a fuller and more detailed account of our key results presented elsewhere \cite{short}.

\section{Thermodynamics: the Yang-Yang Approach}

\subsection{Thermodynamic Bethe Ansatz equations}

The 1D Hubbard model is described by the Hamiltonian
\begin{align}\label{hamiltonian}
H=&-\sum_{j=1}^L \sum_{a=\uparrow,\downarrow} \left( c_{j,a}^\dagger c_{j+1,a} + c_{j+1,a}^\dagger c_{j,a}  \right) \notag\\
&+ u \sum_{j=1}^L \left( 1-2n_{j,\uparrow} \right) \left( 1-2n_{j,\downarrow} \right),
\end{align}
where $c_{j,a}^\dagger$ and $c_{j,a}$ are the creation and annihilation operators of fermions  with spin $a$ ($a=\uparrow$ or $a=\downarrow$) at site $j$ in a
1D  periodic lattice of length $L$, $n_{j,a}=c_{j,a}^\dagger c_{j,a}$ is the corresponding particle number operator,
and $u$ represents an  on-site interaction between particles ($u>0$ for repulsion and $u<0$ for attraction).
By means of the  Bethe ansatz, the eigenenergies of the Hamiltonian are given by $E=-2\sum_{j=1}^N \cos k_j+u(L-2N)$, where
the  quasimomenta  $\left\{k_j\right\}$  satisfy the Lieb-Wu equations \cite{Lieb1968}
\begin{align}
\exp(\mathrm{i} \, k_j  L)&=\prod_{\alpha=1}^M \frac{\sin k_j-\Lambda_\alpha+\mathrm{i} \, u}{\sin k_j-\Lambda_\alpha - \mathrm{i} \, u}, \label{lw1}\\
\prod_{j=1}^N \frac{\sin k_j-\Lambda_\beta + \mathrm{i} \, u}{\sin k_j-\Lambda_\beta - \mathrm{i} \, u}
&=-\prod_{\alpha=1}^M \frac{\Lambda_\alpha-\Lambda_\beta+2 \, \mathrm{i}\,  u}{\Lambda_\alpha-\Lambda_\beta-2 \, \mathrm{i} \, u}, \label{lw2}
\end{align}
where $\left\{\Lambda_\beta \right\}$ denote spin rapidities, $j=1,2,\ldots,N$, $\beta=1,\ldots,M$, with $N$ and $M$ the total particle number and spin down particle number, respectively.

\begin{figure}[h]
\centering
\includegraphics[width=0.4\textwidth]{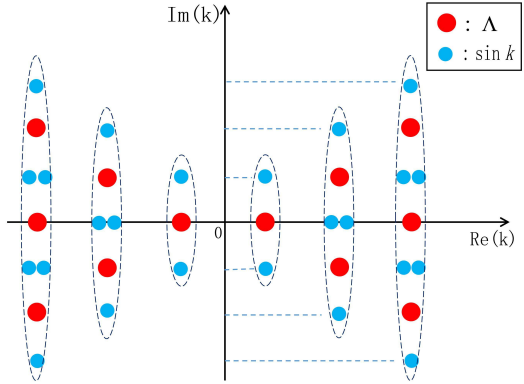}\\
\caption{A schematic configuration  of the $k$-$\Lambda$ strings of length-$1, 2, 3$.
The  $k$-$\Lambda$ bound states are formed by the charge momenta and spin rapidities displayed within the dashed boundaries.
In each $k$-$\Lambda$ bound state  $\sin k$'s  share a real part  with the spin rapidities.
A length-$m$ $k$-$\Lambda$ string contains $m$ rapidities in $\Lambda$-space and $2m$ quasimomenta in $k$-space.
In contrast to the two component Fermi gas,  many electrons on a 1D lattice are allowed to form a bound state of multiparticles.  }
\label{fig:roots}
\end{figure}

Similar to the analysis\cite{MT1972} used for  the repulsive case $u>0$,
one finds that the roots to the Bethe ansatz equations (\ref{lw1}) and (\ref{lw2})
for  the attractive Hubbard model  can be  divided into three categories: single real $k$, $k-\Lambda$ string and $\Lambda-\Lambda$ string,
which constitute the  string hypothesis.
They are given by \cite{Ess1994b,Lee1988}
\begin{itemize}
  \item  single real $k$'s.
 \item  the $\alpha$-th $k$-$\Lambda$ string of length-$\boldsymbol{m}$, for which there are  $\boldsymbol{2m}$ $k$'s,
\begin{align}\label{string-hyp1}
k_\alpha^1&=\arcsin({\Lambda_\alpha^\prime}^m+ \mathrm{i} \, m \, |u|),\nonumber \\
k_\alpha^2&=\arcsin({\Lambda_\alpha^\prime}^m+ \mathrm{i} \,  (m-2) \,  |u|),\nonumber \\
k_\alpha^3&=\pi-k_\alpha^2,\nonumber \\
\vdots \nonumber\\
k_\alpha^{2m-2}&=\arcsin({\Lambda_\alpha^\prime}^m- \mathrm{i} \,(m-2)\, |u|),\nonumber \\
k_\alpha^{2m}&=\arcsin({\Lambda_\alpha^\prime}^m-\mathrm{i} \, m \, |u|),
\end{align}
accompanied by $\boldsymbol{m}$ spin-rapidities
\begin{eqnarray}
{\Lambda_\alpha^\prime}^{m,j}&=&{\Lambda_\alpha^\prime}^m+ \mathrm{i} \, (m+1-2j) \,  |u|,\label{string-hyp2}
\end{eqnarray}
 in  $\Lambda$ space, where  $j=1,2,3,\ldots, m$ and ${\Lambda_\alpha^\prime}^m$ is the real center of the $k-\Lambda$ string, see Fig.~\ref{fig:roots}.
  \item  the $\beta$-th $\Lambda$-$\Lambda$ string of length-$\boldsymbol{m}$,
  \begin{equation}\label{string-hyp3}
  \Lambda_\beta^{m,j}=\Lambda_\beta^m+ \mathrm{i} \, (m+1-2j)\, |u|,
  \end{equation}
where $ j=1,2,3, \ldots, m,$ and $\Lambda_\alpha^m$ is the real center of the $\Lambda$ string.
The $\Lambda$ strings represent the spin wave bound states in the spin sector.
\end{itemize}

In the above equations we  denoted $M_m$, $M_m^\prime$, and $\mathcal{M}_e$ as the number of $\Lambda$ strings of length $\boldsymbol{m}$,
of $k$-$\Lambda$ strings of length-$\boldsymbol{m}$, and of single real $k$'s, respectively.
It is easy to see that $M=\sum_{m=1}^\infty m(M_m+M_m^\prime)$ and $N=\mathcal{M}_e+\sum_{m=1}^\infty 2m M_m^\prime$.

Substituting the string hypothesis into the Lieb-Wu equations and taking logarithms leads to the discrete nested BA equations
\begin{widetext}
\begin{eqnarray}
&&k_j\, L =2\pi\, I_j+\sum_{m=1}^\infty \sum_{\alpha=1}^{M_m^\prime} \theta \left( \frac{\sin k_j-{\Lambda_\alpha^\prime}^m}{m |u|} \right)
+\sum_{m=1}^\infty \sum_{\alpha=1}^{M_m} \theta \left( \frac{\sin k_j-\Lambda_\alpha^m}{m |u|} \right),\\
&&
\sum_{j=1}^{N-2M^\prime} \theta \left( \frac{\Lambda_\alpha^n-\sin k_j}{n |u|} \right)
=2\pi\, J_\alpha^n+\sum_{m=1}^\infty \sum_{\beta=1}^{M_m}\Theta_{n m} \left( \frac{\Lambda_\alpha^n-\Lambda_\beta^m}{n |u|} \right),\\
&&
2L\, \textmd{Re}\left[ \arcsin({\Lambda_\alpha^\prime}^n+\mathrm{i} \, n \, |u|) \right]
=2\pi\, {J_\alpha^\prime}^n+\sum_{j=1}^{N-2M^\prime} \theta \left( \frac{{\Lambda_\alpha^\prime}^n-\sin k_j}{n |u|} \right)
+\sum_{m=1}^\infty  \sum_{\beta=1}^{M_m^\prime}\Theta_{nm} \left( \frac{{\Lambda_\alpha^\prime}^n-{\Lambda_\beta^\prime}^m}{|u|} \right),
\end{eqnarray}
%\end{widetext}
where $M^\prime=\sum_{m=1}^\infty m M_m^\prime$ is the total number of $\Lambda$'s involved in the $k$-$\Lambda$ strings, $\theta(x)=2 \arctan\left(x\right)$, and
%\begin{widetext}
\begin{equation}
\Theta_{nm}(x)=\begin{cases}
\theta \left(\frac{x}{|n-m|} \right) + 2\theta \left( \frac{x}{|n-m|+2} \right)+\cdots+\, 2\theta \left( \frac{x}{n+m-2} \right) + \theta\left( \frac{x}{n+m} \right) & \text{if $n\neq m$}\\
2\theta \left( \frac{x}{2} \right)+2\theta \left( \frac{x}{4} \right)+\cdots+\, 2\theta \left( \frac{x}{2n-2} \right)+\theta \left(\frac{x}{2n} \right) & \text{if $n=m$.}
\end{cases}
\end{equation}
\end{widetext}
The quantum numbers $I_j$, $J_\alpha^n$ and ${J_\alpha^\prime}^n$ are either integers or half-odd integers, stemming from the multivaluedness of
the log functions. They are determined by the  relations
\begin{align}
I_j=&\begin{cases}
\text{integers} & \text{if $\sum_{m=1}^\infty (M_m^\prime+M_m)$ is even}\\
\text{half-odd integers} & \text{if $\sum_{m=1}^\infty (M_m^\prime+M_m)$ is odd,}
\end{cases} \notag\\
J_\alpha^n=&\begin{cases}
\text{integers} & \text{if $N-M_n$ is odd}\\
\text{half-odd integers} & \text{if $N-M_n$ is even,}
\end{cases} \notag\\
{J_\alpha^\prime}^n=&\begin{cases}
\text{integers} & \text{if $L-N+M_n^\prime$ is odd}\\
\text{half-odd integers} & \text{if $L-N+M_n^\prime$ is even.} \notag
\end{cases}
\end{align}

With the help of the string hypothesis, the eigenenergies are
\begin{eqnarray}
E&=&-2\sum_{j=1}^{N-2M^\prime} \cos k_j-2uN+uL \label{energy}\\
&&-4\sum_{n=1}^\infty \sum_{\alpha=1}^{M_n^\prime} \textmd{Re}
\left[ \sqrt{1-({\Lambda_\alpha^\prime}^n+\mathrm{i} \,  n \, |u|)^2} \right].\nonumber
\end{eqnarray}

We now introduce counting functions for the quantum numbers,
$y(k_j)={2\pi\, I_j}/{L}$, $z_n(\Lambda_\alpha^n)={2\pi \, J_\alpha^n}/{L}$ and $z_n^\prime({\Lambda_\alpha^\prime}^n)={2\pi \, {J_\alpha^\prime}^n}/{L}$.
Considering the thermodynamic limit, $N, M, L\to \infty$ with $N/L, \,M/L$ finite, we further define the distributions
\begin{eqnarray}
\frac{\textmd{d}y(k)}{\textmd{d}k}&=&
2\pi \left[\rho^p(k)+\rho^h(k) \right],\nonumber\\
\frac{\textmd{d}z_n(\Lambda)}{\textmd{d}\Lambda}&=&
2\pi \left[\sigma_n^p(\Lambda)+\sigma_n^h(\Lambda)\right],\nonumber\\
\frac{\textmd{d}z_n^\prime(\Lambda)}{\textmd{d}\Lambda}&=&
2\pi \left[{\sigma_n^\prime}^p(\Lambda)+{\sigma_n^\prime}^h(\Lambda)\right],\nonumber
\end{eqnarray}
where $\rho^p$, $\sigma_n^p$, ${\sigma_n^\prime}^p$ ($\rho^h$, $\sigma_n^h$, ${\sigma_n^\prime}^h$) are root densities of particles (holes) in quasimomenta of excess fermions,
$\Lambda$-string parameter space  and $k-\Lambda$ string space, respectively.
Then one can derive the densities of excess fermions, $\Lambda$-spin strings and $k-\Lambda$ strings, with
\begin{eqnarray}
&& \rho^p(k)+\rho^h(k)=\frac{1}{2\pi}\label{rs1}\\
&&-\cos k\sum_{n=1}^\infty\int_{-\infty}^\infty \textmd{d}\Lambda \, a_n(\sin k-\Lambda) \left[ \sigma_n^p(\Lambda)+{\sigma_n^\prime}^p(\Lambda) \right], \nonumber \\
&&\sigma_n^h(\Lambda)= \int_{-\pi}^\pi \textmd{d}k \, a_n(\sin k-\Lambda) \rho^p(k)\nonumber\\
&&-\sum_{m=1}^\infty A_{nm}\ast \sigma_m^p(\Lambda),\label{rs2} \\
&&{\sigma_n^\prime}^h(\Lambda) =\frac{1}{\pi}\textmd{Re} \left[\frac{1}{\sqrt{1-(\Lambda+\mathrm{i} \, n\, |u|)^2}} \right] \label{rs3}\\
&&-\sum_{m=1}^\infty A_{nm}\ast {\sigma_m^\prime}^p (\Lambda) -\int_{-\pi}^\pi \textmd{d}k \, a_n(\sin k-\Lambda)\rho^p(k),\nonumber
\end{eqnarray}
where the function
\begin{equation}
a_n(x)=\frac{1}{2\pi} \frac{2n|u|}{(n|u|)^2+x^2}.\nonumber
\end{equation}
 As usual, $\ast$ stands for the convolution $(f\ast g)(\Lambda)=\int_{-\infty}^{\infty}f(\Lambda-\Lambda')g(\Lambda')d\Lambda'$, namely,
\begin{equation}%\label{Aoperator}
A_{nm}\ast f (x)=\delta_{n, m}\, f(x)+\int_{-\infty}^\infty \frac{\textmd{d}y}{2\pi} \frac{\textmd{d}}{\textmd{d}x}\Theta_{nm}\left( \frac{x-y}{|u|} \right)f(y).\notag
\end{equation}
Here we denoted the derivative of the function  $\Theta_{nm}$ as
\begin{align}
\frac{1}{2\pi}\frac{\texttt{d}}{\texttt{d}x}\Theta_{nm}\left(x\right)
=\begin{cases}
     a_{|n-m|}(x)+2a_{|n-m|+2}(x)\\
     +\ldots+a_{n+m}(x)  & \text{if $n\neq m$} \\
     2a_2(x)+2a_4(x)+\ldots\\
     +2a_{2n-2}(x)+a_{2n}(x) & \text{if $n=m$.}
   \end{cases} \notag
\end{align}

The root  distribution functions (\ref{rs1})-(\ref{rs3}) determine spin and charge excitations, spin dynamics and full energy spectra.
In  the grand canonical ensemble, the Gibbs free energy per site  can be expressed in terms of these root densities in different sectors
\begin{eqnarray}
f&=&\, e-\mu \, n_c-2B \, m-T\, s \nonumber \\
&=& \int_{-\pi}^\pi \textmd{d}k \, (-2\cos k-\mu-2u-B)\rho^p(k)\nonumber\\
&& -\sum_{n=1}^\infty \int_{-\infty}^\infty \textmd{d} \Lambda{\sigma_n^\prime}^p(\Lambda)
\left[ 4\textmd{Re}\sqrt{1-({\Lambda_\alpha^\prime}^n+ \mathrm{i} \, n \, |u|)^2}\right. \nonumber\\
&&\left. + \, n(2\mu+4u) \right] \nonumber \\
&&+\sum_{n=1}^\infty\int_{-\infty}^\infty \textmd{d}\Lambda \, 2n\, B \, \sigma_n^p(\Lambda)-T\, s+u,\label{free2}
\end{eqnarray}
where $\mu$ is the chemical potential, $B$ the magnetic field and $T$ the temperature.
In the above equations $n_c$ is the particle density, $m=\frac{N-2M}{2L}$ the magnetization and $s$ the entropy per site.

Following the Yang-Yang grand can\-onical descrip\-tion \cite{Yang1969}, the entropy per site is explicitly given by
\begin{widetext}
{\small
\begin{eqnarray}
s&=&\int_{-\pi}^\pi \textmd{d}k \, \left\{ \left( \rho^p(k)+\rho^h(k)\right)\ln\left( \rho^p(k)+\rho^h(k)\right) -\rho^p(k)\ln \rho^p(k)-\rho^h(k)\ln \rho^h(k) \right\}\nonumber\\
&& +\sum_{n=1}^\infty \int_{-\infty}^\infty \textmd{d} \Lambda\, \left\{ \left({ \sigma_n^\prime}^p(\Lambda)+{\sigma_n^\prime}^h(\Lambda)\right) \ln \left({ \sigma_n^\prime}^p(\Lambda)+{\sigma_n^\prime}^h(\Lambda)\right) -{\sigma_n^\prime}^p(\Lambda)\ln {\sigma_n^\prime}^p(\Lambda) -{\sigma_n^\prime}^h(\Lambda)\ln  {\sigma_n^\prime}^h(\Lambda)\right\} \nonumber\\
&& +\sum_{n=1}^\infty \int_{-\infty}^\infty \textmd{d}\Lambda \, \left\{ \left( \sigma_n^p(\Lambda)+\sigma_n^h(\Lambda)\right) \ln \left( \sigma_n^p(\Lambda)+\sigma_n^h(\Lambda)\right) -\sigma_n^p(\Lambda)\ln\sigma_n^p(\Lambda) -\sigma_n^h(\Lambda) \ln \sigma_n^h(\Lambda) \right\}.  %\label{enp}
\end{eqnarray}
}
\end{widetext}
In the following, we only consider the physics with $B\geq 0$ and $\mu \leq 0$.

In the  thermodynamic equilibrium, the true equilibrium state can be determined by the minimization of the free energy with respect to  the densities.
Carrying out a  variation of (\ref{free2}) under the restriction  of  (\ref{rs1})-(\ref{rs3}), we obtain  the TBA equations for the attractive Hubbard model
in the form
%\begin{align}
%\ln\zeta(k)=&\frac{-2\cos k-\mu-2u-B}{T}
%+\sum_{n=1}^\infty\int_{-\infty}^\infty \textmd{d}\Lambda \, a_n(\sin k-\Lambda)\ln \left( 1+\frac{1}{\eta_n^\prime(\Lambda)} \right) \nonumber \\
%&-\sum_{n=1}^\infty\int_{-\infty}^\infty \textmd{d}\Lambda \, a_n(\sin k-\Lambda)\ln \left( 1+\frac{1}{\eta_n(\Lambda)} \right), \label{1tba1}\\
%\ln(1+\eta_n(\Lambda))=& \frac{2nB}{T}+\int_{-\pi}^\pi \textmd{d}k \, \cos k  \, a_n(\sin k-\Lambda) \ln \left( 1+\frac{1}{\zeta(k)} \right) \nonumber \\
%&\left. +\sum_{m=1}^\infty A_{nm}\ast \ln\left(1+\frac{1}{\eta_m}\right) \right|_\Lambda, \label{1tba2}\\
%\ln(1+\eta_n^\prime(\Lambda))=&\frac{-4\textmd{Re}\sqrt{1-(\Lambda+i\cdot n \cdot |u|)^2}-n(2\mu+4u)}{T} \nonumber \\
%&+\int_{-\pi}^\pi \textmd{d}k \cos k \, a_n(\sin k-\Lambda)\ln\left(1+\frac{1}{\zeta(k)}\right) \nonumber \\
%&\left. +\sum_{m=1}^\infty A_{nm}\ast \ln\left(1+\frac{1}{\eta_m^\prime}\right)\right|_\Lambda.\label{1tba3}
%\end{align}
%where we have introduced new functions, $\zeta(k)= \rho^h(k)/\rho^p(k)$, $\eta_n(\Lambda)= \sigma_n^h(\Lambda)/\sigma_n^p(\Lambda)$, and
%$\eta_n^\prime (\Lambda)= {\sigma_n^\prime}^h(\Lambda)/{\sigma_n^\prime}^p(\Lambda)$. \cref{1tba1,1tba2,1tba3} are the thermodynamic Bethe ansatz (TBA) equations of the one-dimensional Hubbard model in the attractive regime.
\begin{eqnarray}
\varepsilon^u(k)&=&-2\cos k-\mu-2u-B  \notag\\
&&+\sum_{n=1}^\infty \int_{-\infty}^{\infty} \textmd{d}\Lambda \, a_n(\sin k-\Lambda)\varepsilon_n^{\prime -}(\Lambda)\notag\\
&&-\sum_{n=1}^\infty \int_{-\infty}^{\infty}\textmd{d}\Lambda \, a_n(\sin k-\Lambda)\varepsilon_n^{ -}(\Lambda), \label{tba1} \\
\varepsilon_n(\Lambda)&=&\int_{-\pi}^{\pi}\textmd{d}k \, \cos k \, a_n(\sin k-\Lambda)\varepsilon^{u -}(k)\notag\\
&&+\,2nB+\sum_{m=1}^\infty T_{nm} \ast \varepsilon_m^{ -}(\Lambda),\label{tba2}\\
\varepsilon_n^\prime(\Lambda)&=&-4\textmd{Re}\sqrt{1-(\Lambda+ \mathrm{i} \,  n\, |u|)^2}-n(2\mu+4u)\notag\\
&&+\int_{-\pi}^{\pi}\textmd{d}k \, \cos k \, a_n(\sin k-\Lambda)\varepsilon^{u -}(k)\notag\\
&&+\sum_{m=1}^\infty T_{nm} \ast \varepsilon_m^{\prime -}(\Lambda),\label{tba3}
\end{eqnarray}
where we have denoted
\begin{eqnarray}
\varepsilon^{u -}(x) &=& T\ln\left(1+\mathrm{e}^{-\varepsilon^u(x)/T}\right),\nonumber\\
\varepsilon_n^{\prime -}(x) &=& T\ln\left(1+\mathrm{e}^{-\varepsilon_n^\prime(x)/T}\right),\nonumber\\
\varepsilon_n^{ -}(x) &=&T\ln\left(1+\mathrm{e}^{-\varepsilon_n(x)/T}\right).\nonumber
\end{eqnarray}
In the above equations,  we  defined the dressed energies
\begin{eqnarray}
\varepsilon^u(k)&=&T\ln\zeta(k)=T\ln\rho^h(k)/\rho^p(k),\nonumber\\
\varepsilon_n(\Lambda)&=&T\ln \eta_n(\Lambda)=T\ln \sigma_n^h(\Lambda)/\sigma_n^p(\Lambda),\nonumber\\
\varepsilon_n^\prime(\Lambda)&=&T\ln \eta_n^\prime(\Lambda)=T\ln {\sigma_n^\prime}^h(\Lambda)/{\sigma_n^\prime}^p(\Lambda).\nonumber
\end{eqnarray}
The  convolution $T_{nm}\ast f(x)=A_{nm}\ast f(x)-\delta_{n,m} f(x)$ is defined by convention.

The TBA equations (\ref{tba1})-(\ref{tba3}) indicate that the dressed energies
$\varepsilon^u(k)$, $\varepsilon_n(\Lambda)$, $\varepsilon_n^\prime(\Lambda)$
describe the excitation energies which are subject to interactions among the bound states of electrons, spin wave fluctuations, magnetic field and chemical potential.
 They contain   full  thermal and magnetic fluctuations in both spin and charge degrees of freedom.
 Therefore from these equations we can determine the thermal and magnetic properties of the model in full temperature regimes.
 After some algebra, the Gibbs free energy per site is consequently given by
\begin{eqnarray}
&&f=u-\int_{-\pi}^\pi \frac{\textmd{d}k}{2\pi} \varepsilon^{u -}(k)\label{free3}\\
&&
-\sum_{n=1}^\infty\int_{-\infty}^{\infty}\frac{\textmd{d}\Lambda}{\pi}\textmd{Re}
\left[\frac{1}{\sqrt{1-(\Lambda+ \mathrm{i}  \,  n\, |u|)^2}} \right]\varepsilon_n^{\prime -}(\Lambda).\nonumber
\end{eqnarray}
This result builds up analytical access to the full thermodynamics of the model.

\subsection{Zero Temperature Phase Diagram }

In  the zero temperature limit, most dressed energies are nonnegative and thus make no significant contributions to the free energy (\ref{free3}).
We observe that  in the ground state, there exist only unpaired fermions and bound pairs of fermions.
The spin $\Lambda$-$\Lambda$ strings Eq.(\ref{tba2})  are suppressed due to the fact that in the FFLO-like phase IV, the spin wave bound states ferromagnetically couple to the Fermi sea of  the unpaired fermions.
The driving term in the TBA equation (\ref{tba2}) is positive due to this ferromagnetic ordering. At $T\to 0$, the  $\Lambda$-$\Lambda$ strings  are gapped.
The driving term in the TBA equation (\ref{tba3}) can be positive when $n\ge 2$ due to the negative chemical potential.
Taking the  limit $T\to 0$, the corresponding TBA equations (\ref{tba1}) and (\ref{tba3}) thus reduce to coupled linear integral equations,
called the dressed energy equations,
\begin{align}
\varepsilon^u (k)=&-2\cos k-\mu-2u-B \notag\\
&-\int_{-A}^A \textmd{d}\Lambda \, a_1(\sin k-\Lambda) \varepsilon_1^\prime (\Lambda),\label{0tba1}\\
\varepsilon_1^\prime (\Lambda)=&-2\mu-2\int_{-\pi}^\pi \textmd{d}k \, \cos^2 k\, a_1(\sin k-\Lambda)\notag\\
&-\int_{-Q}^Q \textmd{d}k \, \cos k \, a_1(\sin k-\Lambda) \varepsilon^u (k) \notag \\
&- \int_{-A}^A \textmd{d}\Lambda^\prime \, a_2(\Lambda-\Lambda^\prime)\varepsilon_1^\prime(\Lambda^\prime),\label{0tba2}
\end{align}
where the integration boundaries $Q$ and $A$ represent the Fermi points of these two kinds of states (pairs and single fermions).
In Eq. (\ref{0tba2}) we used the expression
\begin{align*}
&&4\textmd{Re}\sqrt{1-(\Lambda-\mathrm{i}\, n |u|)^2}-4n|u|  \\
&& =\int_{-\pi}^\pi \frac{\textmd{d}k}{\pi} \frac{\cos^2 k \, 2 n |u|}{(nu)^2+(\sin k-
\Lambda)^2}.
\end{align*}
 The integration boundaries  are  determined by  $\varepsilon^{u} (\pm Q)=0$ and $\varepsilon^{\prime}_1(\pm A)=0$.
 Within the intervals $[-Q,Q]$ and $[-A,A]$, the dressed energies  are negative, i.e., $\varepsilon^{u}(k)\leq 0$ and $\varepsilon^{\prime}_1(\Lambda)\leq 0$.
 This means that   particle  states occupy all vacancies in the two Fermi seas.

With the help of (\ref{rs1})-(\ref{rs3}), the root densities for quasimomentum  $k$ and
spin rapidity $\Lambda$ in the $k$-$\Lambda$ string of length-$1$ at  zero temperature are expressed as
\begin{align}
\label{rhok}
\rho(k)=&\frac{1}{2\pi}-\cos k \int_{-A}^A \textmd{d}\Lambda \, a_1(\sin k-\Lambda) \sigma_1^\prime(\Lambda),\\
\sigma_1^\prime(\Lambda)=& \frac{1}{\pi} \textmd{Re}\frac{1}{\sqrt{1-(\Lambda+ \mathrm{i} \,  |u|)^2}}-\int_{-Q}^Q \textmd{d}k \, a_1(\sin k-\Lambda)\rho(k)\notag\\
&-\int_{-A}^A \textmd{d}\Lambda^\prime \, a_2(\Lambda-\Lambda^\prime)\sigma_1^\prime(\Lambda^\prime). \label{sigp1}
\end{align}
In the grand canonical ensemble, we explicitly write down the above root densities,
which  satisfy the two conditions $\int_{-Q}^Q \textmd{d}k \, \rho(k) + 2\int_{-A}^A \textmd{d}\Lambda \, \sigma_1^\prime(\Lambda)=N/L$ and
$\int_{-A}^A \textmd{d}\Lambda \, \sigma_1^\prime (\Lambda)=M/L=N_\downarrow/L$.
Thus the total particle density is given by $n_c = N/L =\int_{-Q}^Q \textmd{d}k \, \rho(k) + 2\int_{-A}^A \textmd{d}\Lambda \, \sigma_1^\prime(\Lambda)$
 and the magnetization per site by $m=(N-2M)/(2L)=\frac{1}{2}\int_{-Q}^Q \textmd{d}k \, \rho(k)$.

By varying the integration boundaries $Q$ and $A$, the system possesses different fillings and quantum phases.
 A phase transition occurs when the dressed energies exactly satisfy $\varepsilon^u(0)=0$, $\varepsilon^u(\pi)=0$ or $\varepsilon_1^\prime(0)=0$.
Consequently we can determine  five phases, (\textmd{I}) vacuum, (\textmd{II}) fully polarized state, (\textmd{III}) half-filling state,
(\textmd{IV}) partially polarized state, i.e., FFLO-like state, (\textmd{V}) fully paired state.
The phase boundary between (\textmd{I}) and (\textmd{V}) is determined by $\varepsilon_1^\prime(0)=0$  together with the condition  $Q=0$.
Then the TBA equation (\ref{0tba2}) leads to the critical field value $\mu_{c}=2|u|-2\sqrt{1+u^2}$.
The phase boundary  between (\textmd{I}) and (\textmd{II}) and between (\textmd{II}) and (\textmd{III})
are determined by the conditions $A=0$, $\varepsilon^u(0)=0$ and by $A=0$, $\varepsilon^u(\pi)=0$, respectively.
With regard to the boundaries for the FFLO-like phase, the situation is much more  subtle.
The phase boundary between (\textmd{II}) and (\textmd{IV}) is determined by $\varepsilon_1^\prime(0)=0$ and $\varepsilon^u(Q)=0$,
while the phase boundary between (\textmd{IV}) and (\textmd{V}) is determined by $\varepsilon^u(0)=0$ and $\varepsilon_1^\prime(A)=0$.

\begin{figure}[h]
\centering
\includegraphics[width=0.5\textwidth]{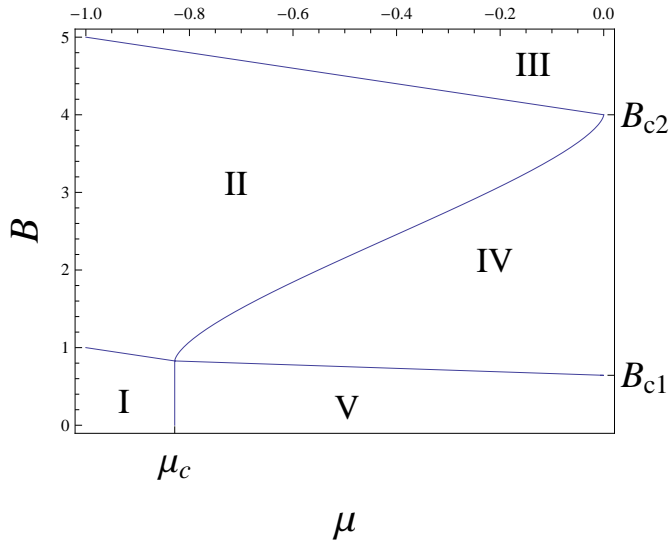}\\
\caption{Ground state phase diagram of  the 1D attractive Hubbard model with $|u|=1$ in the $\mu$-B plane.
In the phase diagram, the critical fields are $\mu_c=2|u|-2\sqrt{1+u^2}<0$,
$B_{c1}=2|u|-2+2\int_{-\infty}^\infty \textmd{d}\omega \, \frac{J_1(\omega) \exp(-|u|\omega)}{w \cosh(u\omega)}$ and
$B_{c2}=2+2|u|$. The different phases are denoted by (\textmd{I}) vacuum, (\textmd{II}) fully polarized state, (\textmd{III}) half-filling state,
(\textmd{IV}) partially polarized state,  (\textmd{V}) fully paired state.
The phase boundaries are defined by equations (\ref{pb15})-(\ref{mu-c5}).
For comparison the low temperature phase diagram is given in Fig.~\ref{wrcomb}. }
\label{phase-diagram}
\end{figure}

The phase boundaries in the ground state phase diagram Fig.~\ref{phase-diagram} are summarized as follows
\begin{itemize}
\item (\textmd{I}-\textmd{V})
\begin{align}\label{pb15}
\mu_{c1}=2|u|-2\sqrt{1+u^2}.
\end{align}
\item (\textmd{I}-\textmd{II})
\begin{align}\label{pb12}
\mu_{c2}=-B-2u-2.
\end{align}
\item (\textmd{II}-\textmd{III})
\begin{align}\label{pb23}
\mu_{c3} = 2-B-2u.
\end{align}
\item (\textmd{II}-\textmd{IV})
\begin{align}
\mu_{c4}=&2|u|-2\sqrt{1+u^2} \notag\\
&-\int_{-Q}^Q \textmd{d}k \, \cos k \, a_1(\sin k)[\cos Q-\cos k],\label{pb24m}\\
B_{c4}=&2\sqrt{1+u^2}-2\cos Q \notag\\
&-\int_{-Q}^Q \textmd{d}k \, \cos k \, a_1(\sin k)[\cos Q-\cos k],\label{pb24b}
\end{align}
with $Q\in [0,\pi]$.
\item (IV-V) This phase transition occurs if the critical magnetic field is sufficient to break the bound state of fermions, whose boundary in principle is fixed by
\begin{align}
\varepsilon_1^\prime (\Lambda)=&-2\mu-2\int_{-\pi}^\pi \textmd{d}k \, \cos^2 k\, a_1(\sin k-\Lambda) \notag\\
&-\int_{-A}^A \textmd{d}\Lambda^\prime \, a_2(\Lambda-\Lambda^\prime)\varepsilon_1^\prime(\Lambda^\prime), \label{tba-pb45}\\
\varepsilon_1^\prime(A)=& \, 0, \label{tba-pb45-A}\\
\mu=& -2-2u-B-\int_{-A}^A \textmd{d}\Lambda \, a_1(\Lambda) \varepsilon_1^\prime (\Lambda).\label{tba-pb45-unpaired}
\end{align}
 When  $A \ll 1$,  the density for pairs of fermions is low, the phase boundary could be obtained  by iteration,
 i.e., by applying Taylor expansion to (\ref{tba-pb45}) with respect to $\Lambda$, it can be approximately resolved by iteration.
The  solution of (\ref{tba-pb45-A}) gives  $A$ in terms of $\mu$ and $B$, then we derive the phase boundary by
substituting the above results for $A$ into the Eqs. (\ref{tba-pb45}) and  (\ref{tba-pb45-unpaired}). By iteration, we finally obtain
\begin{align}
\mu_{c5} \approx & \, 2|u|-B-2 \notag\\
&+\frac{4\sqrt{2}}{\pi |u|\alpha_1}\left[\mu_{c5}+2(\sqrt{1+u^2}-|u|)\right]^{\frac{3}{2}}.\label{mu-c5}
\end{align}
Here at low energy physics only length-$1$ $k$-$\Lambda$ strings  are involved.
From the  TBA equations (\ref{tba1})-(\ref{tba3}),  we may introduce the  parameters $\alpha_n$  and $\beta_n$ to indicate
the interacting effect of the length-$n$ $k$-$\Lambda$ bound states on a lattice in the low density regiem.
They are given by
\begin{eqnarray}
\alpha_n &=&\int_{-\pi}^\pi \textmd{d} k  \,\frac{ 2n\,|u| \cos^2 k (n^2u^2-3 \sin^2 k)}{\pi(n^2u^2+\sin^2 k)^3},\nonumber\\
\beta_n&=& \int_{-\pi}^{\pi} \textmd{d} k \, a_n(\sin k). \nonumber
%\alpha &=&\int_{-\pi}^\pi \textmd{d} k \, \frac{1}{\pi} \,\frac{ 2|u| \cos^2 k (u^2-3 \sin^2 k)}{(u^2+\sin^2 k)^3},\nonumber\\
%\beta &=& \int_{-\pi}^{\pi} \textmd{d} k \, \frac{1}{2\pi}\frac{2|u|}{u^2+\sin^2 k}.\nonumber
\end{eqnarray}
In general, $\alpha_n$ represents the lattice effect in the length-$n$ $k$-$\Lambda$ strings.
\end{itemize}

Meanwhile, if $A \gg 1$, the phase boundary is given  by (\ref{ma-pb45-chem}) and (\ref{ma-pb45-mag})
in Appendix (\ref{app-wienerhopf}), where we have used the Wiener-Hopf method to solve the TBA integral equations.

From the dressed energy equations (\ref{0tba1}) and (\ref{0tba2}) the complete phase diagram at zero temperature is shown in Fig.~\ref{phase-diagram}.
This phase diagram was also obtained by the Shiba transformation, which  builds  up a mapping between repulsive and attractive regions in the ground state
of the Hubbard model \cite{Ess05}.
However, once we are concerned with the low temperature thermodynamics, correlation functions and quantum criticality,
the Shiba transformation does not work in actual calculations,
see the analysis of the ground state properties of the attractive Hubbard model \cite{Ogata1990,Parola1990,Penc1995,Gohmann1998,Ess2000}.
This is mainly because   the different  spin-spin strings, $k-\Lambda$  strings and excess fermions  have different  cut-off  processes (the cut-off strings, see Appendix B) at low temperature physics.
For example, the spin fluctuation term (the third term)  in the unpaired dressed energy can be safely  ignored in the strongly attractive Hubbard model at low temperatures.
However, the counterpart of such a spin  fluctuation term in  the repulsive Hubbard model essentially determines the  antiferromagnetic ordering.
Even in the repulsive regime, such spin string dynamics, quantum criticality and scaling functions still lack an analytical calculation.
The ground state  properties of the attractive Hubbard model were initially studied by Woynarovich \cite{Woyna1991}.
In this paper, using the TBA equations (\ref{tba1})-(\ref{tba3}), we obtain exact results for the FFLO pairing correlation,
universal thermodynamics and quantum criticality of the 1D attractive Hubbard model.
Our study provides a precise  understanding of the universal  low energy physics of interacting fermions with pairing and depairing on  a 1D lattice.

\section{Equation of State}\label{sec-eos}

The TBA equations describe the full  thermodynamics of the model.
At low temperatures quantum liquid behavior and critical scaling in the thermodynamics should be obtained from the TBA equations (\ref{tba1})-(\ref{tba3}).
However, the analysis of such coupled nonlinear integral equations provides a formidable challenge.
In particular, it is challenging to solve infinitely many coupled nonlinear integral TBA equations, i.e., the desired analytical or numerical solution is not achievable by solving the whole set of TBA equations.
 This obstacle prevents us to understand the microscopic Cooper pairing mechanism and many-body phenomena for this model.
 On the other hand,  in the FFLO-like phase IV, the spin wave bound states ferromagnetically couple to the Fermi sea of  the unpaired fermions.
 In this phase, except two gapless excitations in the sectors of bound pairs and excess fermions, there exist a spin wave  ferromagnetic fluctuation, which is no longer a linear dispersion.
 Bosonization or Tomonaga-Luttinger liquid (TLL) theory \cite{Giamarchi:2004} are not available once such a  ferromagnetic ordering is involved in the low temperature physics.
 A similar situation was studied in the 1D two-component Bose gas  \cite{Matveev:2008}.

 Moreover, the TLL  is not  applicable to the quantum critical region near a phase transition.
Here we proceed with an analytical  investigation of the low energy physics of the 1D attractive Hubbard model
beyond the scope of the TLL approaches.
In order to obtain the universal thermodynamics and quantum criticality of the 1D attractive Hubbard model,
we first solve the TBA equations  (\ref{tba1})-(\ref{tba3}) analytically in the strong coupling regime.
We will  derive the equation of state which is crucial for the investigation of the quantum criticality of the model.
These results can be helpful to understand current experimental developments in ultra-cold atoms \cite{Singha2011,Hart2015,Greif2016,Parsons2016,Lawrence2016,Boll2016}.

In the following discussion we  mainly concentrate on the low density regime.
In general, it is very difficult to find universal  characteristics of  quantum liquids in quantum many-body systems,
for example, for  the Gaudin-Yang Fermi gas \cite{YCY2016}.
Under the assumption that the density of  pairs and the bound states of multiple fermions are low and the interaction is strong,
the TBA equations (\ref{tba1})-(\ref{tba3})  can be rewritten as
\begin{align}
\varepsilon^u(k)=& -2\cos k + 2\bar{a} \cos^2 k - \mu -2u -B +\sum_{n=1}^\infty p_n^b \notag\\
&+ \bar{a} - T e^{-2B/T} e^{-\bar{K}} I_0(\bar{K})+o\left(\frac{1}{|u|^4} \right),\label{reduced-tba-u}\\
\varepsilon_n^\prime(\Lambda)=&-2n\mu+\eta_n-\frac{d_1}{\pi n |u|} - \frac{d_2}{\pi (n|u|)^3} \notag\\
&+\Lambda^2 \left[ \frac{d_1}{\pi(n|u|)^3}-\varphi_n \right] +o\left(\frac{1}{|u|^4} \right),\label{reduced-tba-b}
\end{align}
%\begin{align}
%\varepsilon_n^\prime(\Lambda)=&-2n\mu-4\left(nu+\sqrt{(nu)^2+1} \right) + K_n^1 +T\sum_{m=1}^\infty \int_{-\infty}^\infty \textmd{d}\Lambda \, T_{nm}(\Lambda) \ln \left(1+e^{\varepsilon_m^\prime(\Lambda)/T} \right) \notag\\
%&+ \mathfrak{L}_n \Lambda^2 + K_n^2 \Lambda^2 -\Lambda^2 T\sum_{m=1}^\infty \int_{-\infty}^\infty \textmd{d}\Lambda \, Q_{nm}(\Lambda) \ln %\left(1+e^{\varepsilon_m^\prime(\Lambda)/T} \right)
%\end{align}
where $\bar{K}=\int_{-\pi}^\pi  \frac{\textmd{d}k}{2\pi} \, \cos k \, \ln(1+\mathrm{e}^{-\varepsilon^u(k)/T})$
and $I_0(x)$ is the zeroth order modified Bessel function, which stems  from the spin-wave contributions.
In the above equations, we denoted
\begin{eqnarray}
d_1&=&2\pi-\int_{-\pi}^\pi \textmd{d}k \, \cos k \,\varepsilon^{u -}(k),\nonumber\\
 d_2&=&-\frac{\pi}{2}-\int_{-\pi}^\pi \textmd{d}k \, \cos k \, \sin^2 k \, \varepsilon^{ u-}(k),\nonumber \\
 \eta_n&=&\sum_{m=1}^\infty \int_{-\infty}^\infty \textmd{d}\Lambda \, T_{nm}(\Lambda)\varepsilon_m^{ \prime -}(\Lambda),\nonumber\\
  \bar{a}&=&\frac{1}{2} \sum_{n=1}^\infty \int_{-\infty}^\infty \textmd{d}\Lambda \left[b_n(\Lambda)-\frac{4 b_n(\Lambda)\Lambda^2}{(n u)^2+\Lambda^2} \right]
  \varepsilon_n^{\prime  -}(\Lambda),\nonumber\\
 \varphi_n&=&\sum_{m=1}^\infty \int_{-\infty}^\infty \textmd{d}\Lambda \, Q_{nm}(\Lambda)\varepsilon_m^{\prime  -}(\Lambda), \nonumber
 \end{eqnarray}
 with  $b_n(\Lambda)=\frac{a_n(\Lambda)}{(n u)^2+\Lambda^2}$ and
\begin{align}
Q_{nm}(x)=
\begin{cases}
b_{|n-m|}(x)+2b_{|n-m|+2}(x)+\ldots\\
+2b_{n+m-2}(x)+b_{n+m}(x) \quad \text{if $n \neq m$} \\
2b_2(x)+2b_4(x)+\ldots \\
+2b_{2n-2}(x)+b_{2n}(x) \qquad \text{if $n=m$.}
\end{cases}
\end{align}
The results (\ref{reduced-tba-u}) and (\ref{reduced-tba-b}) are valid for the low density limit and strong interaction regime.

Substituting \cref{reduced-tba-u,reduced-tba-b} into (\ref{free3}), the pressure per unit length is given by  $p=p^u+\sum_{n=1}^\infty p_n^b+|u|$ with
\begin{widetext}
\begin{equation}
p^u=T\int_{-\pi}^\pi \frac{\textmd{d}k}{2\pi} \ln\left(1+ \mathrm{e}^{-\varepsilon^u(k)/T}\right), \quad
p_n^b=T\int_{-\infty}^{\infty}\frac{\textmd{d}\Lambda}{\pi}\textmd{Re}
\left[\frac{1}{\sqrt{1-(\Lambda+ \mathrm{i} \, n\, |u|)^2}} \right]\ln\left(1+\mathrm{e}^{-\frac{\varepsilon_n^\prime(\Lambda)}{T}}\right).  \label{pb-a}
\end{equation}
\end{widetext}
Using the results (\ref{reduced-tba-u}) and (\ref{reduced-tba-b}) and taking integration by parts within the above expressions for the effective pressures (\ref{pb-a}),
we then obtain the set of coupled equations
\begin{eqnarray}
p^u&=&\, T \ln \left( 1+\mathrm{e}^{(\mu+2u+B-\sum_{n=1}^\infty p_n^b + \bar{a} -2 )/T} \right)  \label{pressure-a1}\\
&& -\frac{2\bar{a}}{\pi} \int_{-1}^1 \textmd{d}x \, \frac{x^2/\sqrt{1-x^2}}{1+\mathrm{e}^{2x/T}/z} +\frac{2\bar{a}}{1+\mathrm{e}^{4/T}/z^2}\notag\\
& &+\frac{2}{\pi} \int_{-1}^1 \textmd{d}x \, \frac{\arccos(-x)}{1+\mathrm{e}^{2x/T}/z} +o\left(\frac{1}{u^4}\right), \notag\\
p_n^b&=&\, T\left[1-\frac{1}{4(nu)^2}\right]\ln\left(1+\mathrm{e}^{2n\mu/T}\right) \label{pressure-a2}\\
&&+\frac{2 D_n}{\pi} \left[1-\frac{1}{4(nu)^2} \right] \int_0^\infty \textmd{d}x \, \frac{\arctan\sqrt{x}}{1+\mathrm{e}^{D_n x/T}/\zeta_n} \nonumber \\
&&+o\left(\frac{1}{u^4}\right), \nonumber
\end{eqnarray}
 which serve as the equations of state.
In the above equations,
\begin{eqnarray}
D_n&=&\frac{d_1}{\pi n|u|}-(n u)^2\varphi_n,\nonumber\\
z&=&\mathrm{e}^{\left( \mu+2u+B-\sum_{n=1}^\infty p_n^b +\bar{a} \right)/T},\nonumber\\
\zeta_n&=&\mathrm{e}^{\left( 2n\mu-\eta_n+\frac{d_1}{\pi n |u|}+\frac{d_2}{\pi(n|u|)^3} \right) /T}.\nonumber
\end{eqnarray}

We also  defined  the auxiliary functions
\begin{eqnarray}
d_1&=&\,2\pi-4 \int_{-1}^1 \textmd{d}x \, \frac{\sqrt{1-x^2}}{1+\mathrm{e}^{2x/T}/z} \notag\\
&&- 4\bar{a} \int_{-1}^1 \textmd{d}x \, \frac{x^3/\sqrt{1-x^2}}{1+\mathrm{e}^{2x/T}/z}+o\left(\frac{1}{u^4}\right),\notag\\
d_2&=&\, -\frac{\pi}{2} - \frac{4}{3} \int_{-1}^1 \textmd{d}x \, \frac{(1-x^2)^{3/2}}{1+\mathrm{e}^{2x/T}/z}+o\left(\frac{1}{u^4}\right),\notag\\
\bar{a}&=&\, \sum_{n=1}^\infty \frac{D_n}{\pi (n u)^2} \int_0^\infty \textmd{d}x \, \frac{\sqrt{x}/(1+x)^2}{1+\mathrm{e}^{D_n x/T}/\zeta_n}+o\left(\frac{1}{u^4}\right),\notag\\
\eta_n&=& \sum_{m=1}^\infty \mathfrak{T}_{nm}^\xi (m)+o\left(\frac{1}{u^4}\right), \notag\\
\varphi_n&=& \sum_{m=1}^\infty \mathfrak{T}_{nm}^\phi (m)+o\left(\frac{1}{u^6}\right).
\end{eqnarray}
In these equations, we define $\mathfrak{T}_{nm}^x (m)= x_{|n-m|}^m+2x_{|n-m|+2}^m + \cdots + 2x_{n+m-2}^m + x_{n+m}^m$, with $x_0^m=0$ ($x=\eta,\phi$) and  auxiliary functions
\begin{align}
\xi_p^m=& \, T\ln\left(1+\mathrm{e}^{2m\mu/T}\right) \notag\\
&+ \frac{2 D_m}{\pi} \int_0^\infty \textmd{d}x \, \frac{\arctan\left( \frac{m}{p}\sqrt{x} \right)}{1+\mathrm{e}^{D_m x/T}/\zeta_m},\notag\\
\phi_p^m=& \,\frac{T}{2(p u)^2} \left(1+\mathrm{e}^{2m\mu/T}\right) \notag\\
&+ \frac{m}{p} \frac{D_m}{\pi u^2} \int_0^\infty \textmd{d}x \, \frac{\sqrt{x}/(p^2+m^2 x)}{1+\mathrm{e}^{D_m x/T}/\zeta_m} \notag\\
&+ \frac{D_m}{\pi (p u)^2} \int_0^\infty \textmd{d}x \,  \frac{\arctan \left( \frac{m}{p} \sqrt{x} \right)}{1+\mathrm{e}^{D_m x/T}/\zeta_m}.
\end{align}
These functions are indicative of the  sophisticated many-body effects induced by $k$-$\Lambda$ strings of different lengths.
A more detailed derivation of the above result is presented in Appendix \ref{app-eos}.

\begin{figure}[!htb]
\centering
\includegraphics[width = 0.45\textwidth]{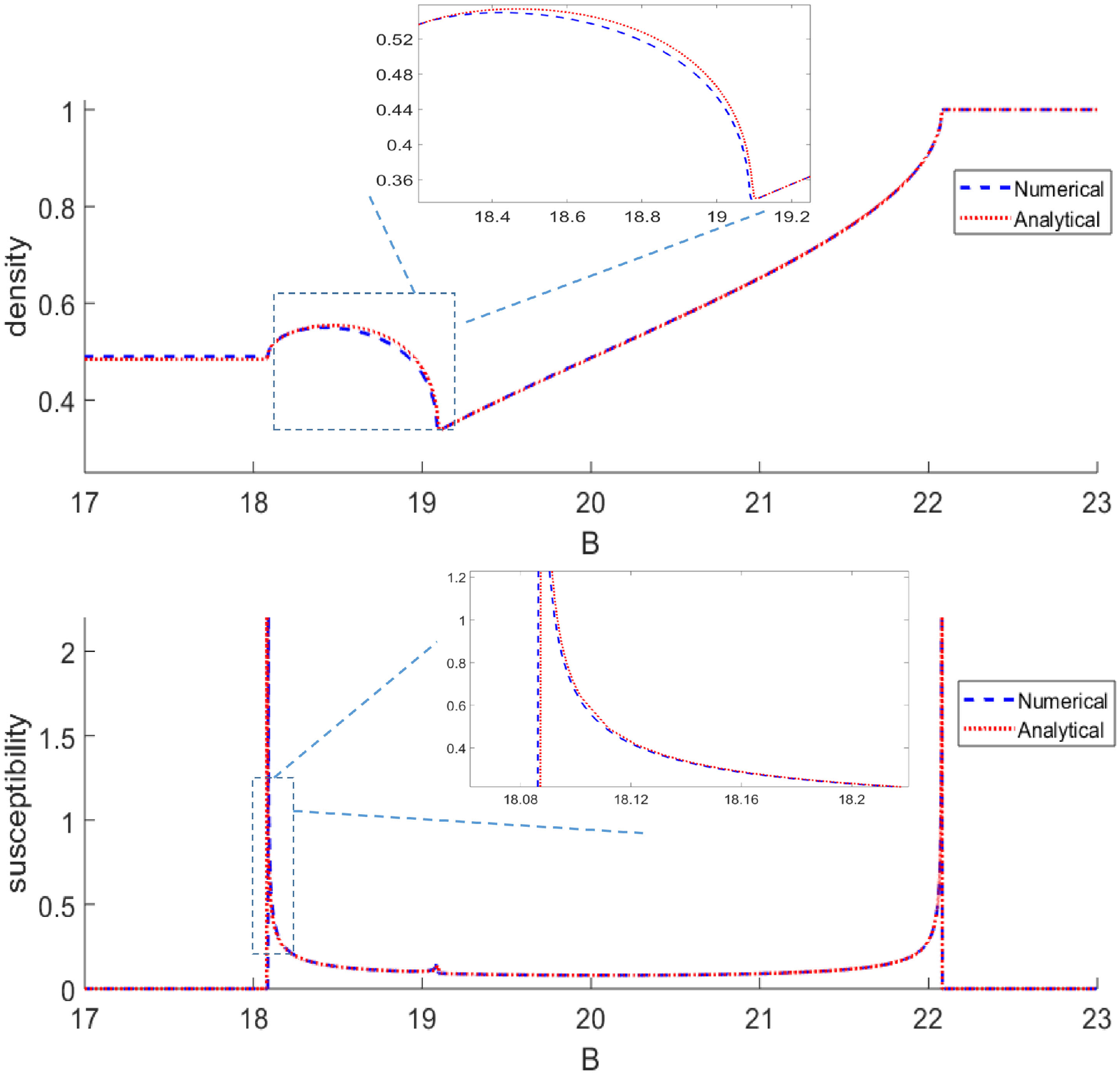}
\caption{A comparison between the analytic results (\ref{eos-pu}) and (\ref{eos-pb}) and the numerical results obtained from the TBA equations (\ref{tba1})-(\ref{tba3}).
We set up natural units in the plots. The upper and the lower panels respectively show the density and susceptibility vs. magnetic field across phases V, IV, II, III
at a fixed chemical potential $\mu=-0.08$,  temperature $T=10^{-4}$ and interaction strength $u=-10$. The sudden changes in the density and susceptibility
show subtle scaling behavior near  phase transitions.}
\label{demon}
\end{figure}

In order to conceive the universal behavior of the system, we need to further simplify the equations of state (\ref{pressure-a1}) and  (\ref{pressure-a2}).
 To this end, we utilize the conditions $|\frac{\mu}{T}| \gg 1$ and strong interaction $|u|\gg 1$, which suppress the large length  $k$-$\Lambda$ strings in  this physical regime.
 We observe  that  no larger length-$n$  $k$-$\Lambda$ bound states than $n=1$ exist in the FFLO  phase IV at low temperatures.
 Then the pressure per unit length  simplifies to $p= p^u + p^b + |u|$, where  $p^u$ and $p^b$ are given by
\begin{align}
p^u=&\, T\ln\left(1+\mathrm{e}^{(\mu+2u+B-p^b-2)/T}\right) \notag\\
&+\frac{2}{\pi} \int_{-1}^1 \textmd{d}x \, \frac{\arccos(-x)}{1+\mathrm{e}^{2x/T}/z_1}+o\left(\frac{1}{u^2}\right), \label{eos-pu} \\
p^b=&\, \frac{2 D_1}{\pi} \int_0^\infty \textmd{d}x \, \frac{\arctan\sqrt{x}}{1+\mathrm{e}^{D_1 x/T}/\zeta}+o\left(\frac{1}{u^2}\right), \label{eos-pb}
\end{align}
where $z_1=\mathrm{e}^{\left( \mu+2u+B-p^b \right)/T}$, $\zeta=\mathrm{e}^{\left( 2\mu-\eta+\frac{d_1}{\pi |u|} \right)/T}$
and the above  auxiliary functions with $n=1$ read
\begin{eqnarray}
D_1&=&\frac{d_1}{\pi |u|}-u^2\varphi,\\
d_1&=&\,2\pi-4 \int_{-1}^1 \textmd{d}x \, \frac{\sqrt{1-x^2}}{1+\mathrm{e}^{2x/T}/z_1}+o\left(\frac{1}{u^2}\right),\label{eos-d1}\\
\eta&=&\, \frac{2 D_1}{\pi} \int_0^\infty \textmd{d}x \, \frac{\arctan\left( \frac{1}{2}\sqrt{x} \right)}{1+\mathrm{e}^{D_1 x/T}/\zeta}+o\left(\frac{1}{u^2}\right), \label{eos-eta}\\
\varphi&=&\frac{D_1}{2\pi u^2} \int_0^\infty \textmd{d}x \, \frac{\sqrt{x}/(4+x)}{1+\mathrm{e}^{D_1 x/T}/\zeta} \notag\\
&&+ \frac{D_1}{4\pi u^2} \int_0^\infty \textmd{d}x \,  \frac{\arctan \left( \frac{1}{2} \sqrt{x} \right)}{1+\mathrm{e}^{D_1 x/T}/\zeta}+o\left(\frac{1}{u^4}\right). \label{eos-phi}
\end{eqnarray}
Here we only consider the corrections up to order ${1}/{|u|}$ in the strong coupling regime $|u|\gg 1$.
The equations of state (\ref{eos-pu}) and (\ref{eos-pb})  give a very good approximation of the low energy physics.
In Fig.~\ref{demon}, we demonstrate the  accuracy of these  equations  compared to the numerical results obtained from the TBA equations (\ref{tba1})-(\ref{tba3}).
The peaks in the susceptibility and the discontinuities of the first derivative  of  the density reveal important behavior of the model near quantum phase transitions.

The pressures (\ref{eos-pu}) and (\ref{eos-pb}) could be further approximately resolved by appropriate iteration.
For the low density regime $n_c\ll 1$,  we  expand the numerators  in the pressure $p^b$
and the auxiliary functions $\eta$ and $\varphi$  with respect to a small value of   $x$ in these integrals.
Then we can represent $\eta$ and $\varphi$ in terms of $p^b$.
After iteration we  thus obtain  $p^b\approx -T^{\frac{3}{2}} \, f_{\frac{3}{2}}/\sqrt{\frac{d_1}{|u|}-\frac{\pi p^b}{8}}$,
where we have defined $f_s=\textmd{Li}_s \left[-\exp\left( \frac{1}{T} \left( 2\mu-\frac{p^b}{2}+\frac{d_1}{\pi |u|} \right) \right)\right]$
in terms of the polylog function $\textmd{Li}_s\left(x\right)$.
Using this expression for $p^b$ and after some lengthy algebra,  we finally obtain the closed form expressions
\begin{align}
p^b=&-\frac{1}{\sqrt{\pi D_0}} \, T^{\frac{3}{2}} \tilde{f}_{\frac{3}{2}}+o\left(\frac{1}{u^2},T^2 \right),\label{effective-pb}\\
p^u=&\,T\ln\left( 1+\mathrm{e}^{(\mu+2u+B- p^b-2)/T} \right)\label{effective-pu} \\
&+\frac{2}{\pi}\int_{-1}^1 \textmd{d}x \, \frac{\arccos(-x)}{1+\mathrm{e}^{2x/T}/(z_0 \, \mathrm{e}^{-p^b/T})}+o\left(\frac{1}{u^2},T^2 \right)\nonumber
\end{align}
for the two pressures, with the auxiliary functions
\begin{align}
d_0=&2\pi-4 \int_{-1}^1 \textmd{d}x \, \frac{\sqrt{1-x^2}}{1+\mathrm{e}^{2x/T}/z_0}+o\left(\frac{1}{u^2},T^2 \right),\label{effective-d0}\\
D_0=&\frac{d_0}{|u|\pi}+\frac{1}{8}\sqrt{\frac{|u|}{d_0}} \, T^{\frac{3}{2}}g_{\frac{3}{2}}+o\left(\frac{1}{u^2},T^2 \right).\label{effective-D1}
\end{align}
In results (\ref{effective-pb}) and (\ref{effective-pu}),
$\tilde{f}_s=g_s-\frac{1}{2}\sqrt{\frac{|u|}{d_0}} \, T^{\frac{1}{2}} g_s\, g_{s-1}
$, $z_0=\mathrm{e}^{\left(\mu+2u+B\right)/T}$
and $  g_s=\textrm{Li}_s\left[ -\mathrm{e}^{\left(2\mu+\frac{d_0}{|u|\pi}\right)/T} \right]$.
The pressures (\ref{effective-pb}) and (\ref{effective-pu}) give deep insight into quantum scaling in the critical regimes.

\section{Quantum Criticality}\label{sec-qc}

 Quantum phase transitions occur in the attractive Hubbard model at zero temperature as the external magnetic field and chemical potential
 are varied across any phase boundary in Fig.~\ref{phase-diagram}.
In general, near a quantum critical point, the model  is expected to show universal scaling behaviour in the thermodynamic quantities due to the collective nature of many-body effects \cite{Fisher1989}.
We see that the 1D attractive Hubbard model is an ideal model to explore such a universal  scale-invariant description on a 1D lattice,
which  can be determined by the power-law scaling of the various thermodynamic properties.
The behavior of the thermodynamic quantities  is governed by scaling functions with critical exponents
in the V-shaped region fanning out to  finite temperatures from the quantum critical point.
 In order to calculate the thermodynamic quantities which contain enough thermal and quantum fluctuations to describe quantum criticality,
 we here use the form of the equation of state with the results given in (\ref{eos-pu}), (\ref{eos-pb}), and (\ref{eos-d1})-(\ref{eos-phi}) for the pressure terms.
 We observe that  first-order derivatives of these pressures  with respect to $\mu$ or $B$ form a set of linear equations.
 Solution to this set of linear equations directly leads  to the particle density $n_c=\left(\frac{\partial p}{\partial \mu}\right)_B$ and magnetization $m=\frac{1}{2}\left(\frac{\partial p}{\partial B}\right)_\mu$.
 Similarly, one can  derive the second-order derivatives of the pressures, the compressibility $\kappa=\left(\frac{\partial n}{\partial \mu}\right)_B$ and  the susceptibility $\chi=\left(\frac{\partial m}{\partial B}\right)_\mu$.
The corresponding scaling laws can be obtained in the different physical regimes.

\begin{figure}[ht]
\includegraphics[width=0.50\textwidth]{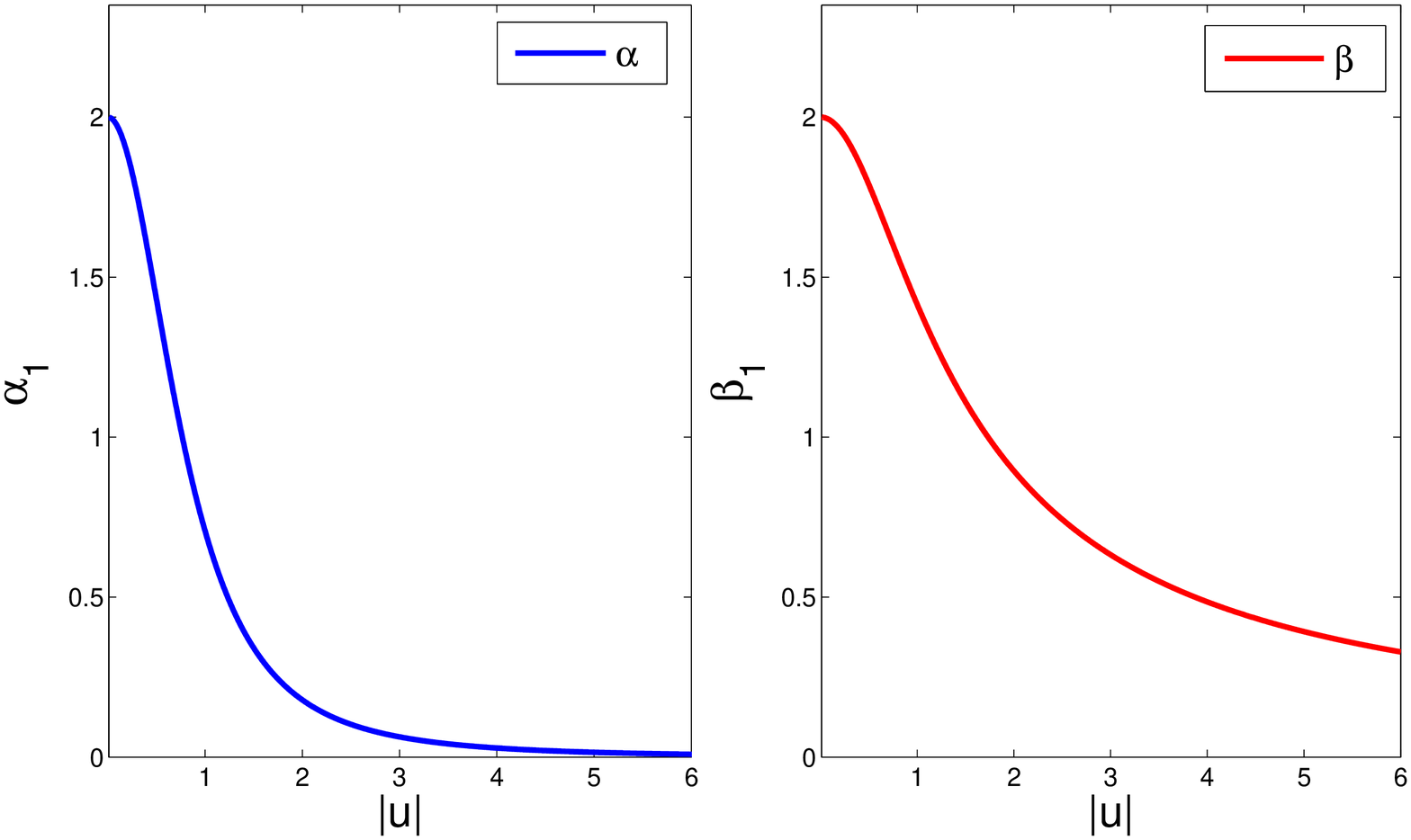}
\caption{The lattice interacting parameters for the length-$1$ $k-\Lambda$ strings as a function of the interaction strength $u$.
The  parameter $\alpha_1$ strongly affects the  band dispersion of bound pairs.
The paramter $\beta_1$ presents a lattice contribution to the free energy of the pairs.
}
\label{Fig:lattice}
\end{figure}

\begin{figure}[!htb]
\centering
\includegraphics[width = 0.5\textwidth]{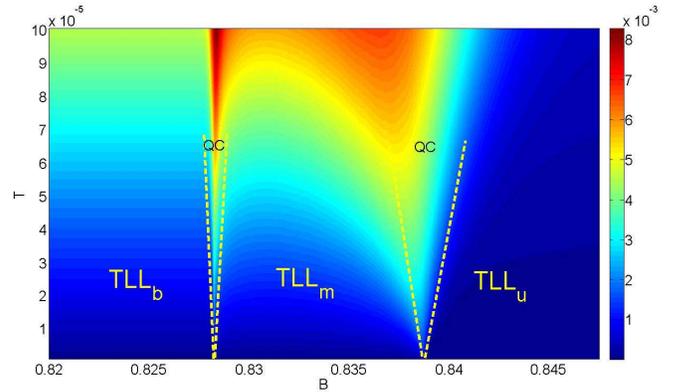}
\caption{Contour plot entropy  vs. magnetic field $B$ for the 1D attractive Hubbard model.
The numerical calculation is performed by solving  the TBA equations (\ref{tba1})-(\ref{tba3})
with  a fixed  chemical potential $\mu=-0.828$ and interaction $u=-1$.
The crossover temperatures (white dashed lines) fanning out from the critical points separate
different TLL  phases from the  quantum critical regimes.
The linear temperature-dependent entropy breaks down when the temperature is greater than these crossover temperatures.
Here  $\textrm{TLL}_\textrm{u}$ and $\textrm{TLL}_\textrm{b}$ respectively stand for the TLLs of unpaired fermions and  bound pairs.
$\textrm{TLL}_\textrm{m}$ stands for the two-component TLL of the FFLO-like state. }
\label{entropy}
\end{figure}

At very low temperatures, spin fluctuation in the FFLO-like phase is suppressed,
as are the bound states of higher $k$-$\Lambda$ strings for $|{\mu}/{T}| \gg 1$.
In this regime, the thermodynamics of the model  is governed by a two-component TLL
or say two-component Fermi liquid consisting   of excess fermions and of hard-core bosonic  charge bound states.
 The leading low-temperature correction to the free energy is given by
\begin{align}
f \approx f_0 - \frac{\pi T^2}{6} \left( \frac{1}{v_1} + \frac{1}{v_2} \right), \label{low-tem-correction}
\end{align}
where $f_0$ is the ground state free energy and $v_1$ ($v_2$) is the sound velocity of excess fermions (bound pairs).
This result is valid for arbitrary interaction strength.
When  the particle density is very low, i.e., $n_{1,2} \ll 1$,  we   explicitly obtain the two velocities
\begin{align}
v_1\approx & \, 2\pi n_1 \left[1+4\,\frac{n_2}{|u|}+12\left(\frac{n_2}{|u|}\right)^2  \right], \notag\\
v_2\approx & \, \pi \, n_2 \frac{\sqrt{2\alpha_1}}{\beta_1} \left[1+\frac{1}{\beta_1}\left(2\,\frac{n_1}{|u|}+\frac{n_2}{|u|}\right) \right. \notag\\
&\left. + \frac{3}{\beta_1^2}\left(2\,\frac{n_1}{|u|}+\frac{n_2}{|u|} \right)^2 \right],\label{velocity}
\end{align}
where the lattice  parameters
\begin{eqnarray}
\alpha_1&=&\int_{-\pi}^\pi \textmd{d}k \, \frac{2|u|\cos^2 k \, (u^2-3\sin^2 k)}{\pi(\sin^2 k+ u^2)^3},\nonumber\\
\beta_1&=&\int_{-\pi}^\pi \textmd{d}k\, a_1(\sin k),
 \end{eqnarray}
 are  functions of $|u|$ representing the lattice effect \cite{note1}.
 In the above equations, $n_{1,2}$ stands for the densities of excess fermions and the bound pairs, respectively.
We plot the two lattice parameters against interaction strength in Fig.~\ref{Fig:lattice}.
We shall see that the  critical exponents and thermodynamics of the model are subject to these two   parameters.
The susceptibility is independent of temperature so that the dimensionless Wilson ratio reaches a constant (we will study this nature of the Fermi liquid in the next section).
The TLL validates only in the region below the crossover temperatures, where the entropy or specific heat retains a linear temperature-dependence, see the dashed lines in Fig.~\ref{entropy}.
The entropy in the temperature-magnetic field plane displays the visible areas of the critical regions (QC) near different critical points.
In what follows, we will derive the scaling functions for the critical regions.

Using the equation of state with the pressures (\ref{eos-pu}), (\ref{eos-pb}), and (\ref{eos-d1})-(\ref{eos-phi}),
we can further derive the scaling forms of the thermodynamic quantities in the  critical regimes.
Analytic results for the scaling functions help to understand the  microscopic origin of quantum criticality of  the 1D attractive Hubbard model.
For convenience, we first simplify the auxiliary functions
\begin{align}
\delta=& \frac{1}{\pi}\int_{-1}^1 \textmd{d}x \, \frac{1}{1+\mathrm{e}^{2x/T}/\bar{z}_0} \frac{x}{\sqrt{1-x^2}},\notag\\
\gamma=& \frac{1}{\pi} \int_{-1}^1 \textmd{d}x \, \frac{1}{1+\mathrm{e}^{2x/T}/\bar{z}_0} \frac{1}{\sqrt{1-x^2}},\notag\\
\gamma'=& \frac{1}{\pi T} \int_{-1}^1 \textmd{d}x \, \frac{e^{2x/T}/\bar{z}_0}{(1+\mathrm{e}^{2x/T}/\bar{z}_0)^2} \frac{1}{\sqrt{1-x^2}},\notag\\
\delta'=& \frac{1}{\pi T} \int_{-1}^1 \textmd{d}x \, \frac{e^{2x/T}/\bar{z}_0}{(1+\mathrm{e}^{2x/T}/\bar{z}_0)^2} \frac{x}{\sqrt{1-x^2}},
\end{align}
where $\bar{z}_0=\exp\left( \mu+2u+B+ T^{\frac{3}{2}} g_{\frac{3}{2}} / \sqrt{\pi D_0} \right)$ with $D_0$ given in (\ref{effective-D1}).
By virtue of results (\ref{eos-pu}), (\ref{eos-pb}), and (\ref{eos-d1})-(\ref{eos-phi}), the closed form expressions for thermodynamic quantities
can be derived.
For strong attraction, we have the relations
\begin{align}
{n}_c=&\,\gamma+(1-\gamma)\frac{\partial p^b}{\partial \mu},\label{TD-n}\\
{m}=&\,\frac{1}{2}\left[ \gamma+(1-\gamma)\frac{\partial p^b}{\partial B}  \right],\label{TD-m}\\
{\kappa}=&\,\left(1- \frac{\partial p^b}{\partial \mu} \right)^2 \gamma' + (1-\gamma) \frac{\partial^2 p^b}{\partial \mu^2},\label{TD-kappa}\\
{\chi}=&\, \frac{1}{2}\left[ \left(1- \frac{\partial p^b}{\partial B} \right)^2 \gamma' + (1-\gamma) \frac{\partial^2 p^b}{\partial B^2} \right],\label{sus}
\end{align}
for thermodynamic quantities. Here we calculated the derivatives of the pressures
%\begin{align}
%\frac{\partial p^b}{\partial \mu}=& -\frac{8\, \tau^{\frac{1}{2}} \left(2 f_{\frac{1}{2}}-\tau f_{\frac{3}{2}}\right)}{\Delta_t} ,\\
%\frac{\partial p^b}{\partial B}=& -\frac{16\, \delta \tau^{\frac{1}{2}} \left(f_{\frac{1}{2}} - \tau f_{\frac{3}{2}} \right)}{|u|\Delta_t} ,\\
%\frac{\partial^2 p^b}{\partial \mu^2}=& -\frac{128 f_{-\frac{1}{2}} \left(16 \pi - \sqrt{\pi} \tau^{\frac{3}{2}} f_{\frac{3}{2}}\right)}{D_0 \tau^{\frac{1}{2}} \Delta_t^3},\\
%\frac{\partial^2 p^b}{\partial B^2}=& -\frac{16\delta' \tau^{\frac{1}{2}}(f_{\frac{1}{2}}-\tau f_{\frac{3}{2}})}{|u|\Delta_t}
%-\frac{4096 \pi \delta^2 f_{-\frac{1}{2}}}{u^2 D_1 \tau^{\frac{1}{2}} \left(4 f_{\frac{1}{2}}-\tau f_{\frac{3}{2}}\right) \Delta _t^4} \notag\\
%&\times \left( 16 \sqrt{\pi}f_{\frac{1}{2}} -8 \tau^{\frac{1}{2}} f_{\frac{1}{2}}^2 -5  \tau ^{\frac{3}{2}} f_{\frac{1}{2}} f_{\frac{3}{2}} -4 \sqrt{\pi} \tau  f_{\frac{3}{2}}\right)
%\end{align}
\begin{eqnarray}
\frac{\partial p^b}{\partial \mu}&=& -\frac{\tau^{\frac{1}{2}} \left(2 f_{\frac{1}{2}}-\tau f_{\frac{3}{2}}\right)}{\Delta_t} ,\\
\frac{\partial p^b}{\partial B}&=& -\frac{2\, \delta \tau^{\frac{1}{2}} \left(f_{\frac{1}{2}} - \tau f_{\frac{3}{2}} \right)}{|u|\Delta_t} ,\\
\frac{\partial^2 p^b}{\partial \mu^2}&=& -\frac{ f_{-\frac{1}{2}} \left(16 \pi - \sqrt{\pi} \tau^{\frac{3}{2}} f_{\frac{3}{2}}\right)}{4D_0 \tau^{\frac{1}{2}} \Delta_t^3},\\
\frac{\partial^2 p^b}{\partial B^2}&=& -\frac{2\delta' \tau^{\frac{1}{2}}(f_{\frac{1}{2}}-\tau f_{\frac{3}{2}})}{|u|\Delta_t}
-\frac{\pi \delta^2 f_{-\frac{1}{2}}}{u^2 D_1 \tau^{\frac{1}{2}} \left(4 f_{\frac{1}{2}}-\tau f_{\frac{3}{2}}\right) \Delta _t^4} \notag\\
&&\times \left( 16 \sqrt{\pi}f_{\frac{1}{2}} -8 \tau^{\frac{1}{2}} f_{\frac{1}{2}}^2 -5  \tau ^{\frac{3}{2}} f_{\frac{1}{2}} f_{\frac{3}{2}} -4 \sqrt{\pi} \tau  f_{\frac{3}{2}}\right)
\end{eqnarray}
with
$
\tau=T/D_0
$ and
$
\Delta_t=\sqrt{\pi }- \frac{1}{2}\tau^{\frac{1}{2}} f_{\frac{1}{2}}+\frac{1}{8}\tau^{\frac{3}{2}} f_{\frac{3}{2}}
$.
These results constitute very accurate results for the thermodynamics.
The asymptotic results for the thermodynamic properties (\ref{TD-n})-(\ref{sus}) have been demonstrated in Fig.~\ref{demon}.

%\begin{figure}[!htb]
%\centering
%\subfloat[]{\label{demon-den}\includegraphics[width = 0.45\linewidth]{demon-den.eps}}
%\subfloat[]{\label{demon-sus}\includegraphics[width = 0.45\linewidth]{demon-sus.eps}}\\
%\caption{A demonstration between the analytic results and the numerical ones is made, where \cref{demon-den} and \cref{demon-sus} show the density and susceptibility vs magnetic field across \textit{Phase V, IV, II, III} with fixed $\mu=-0.07$, $T=10^{-3}$K and $u=-10$.}
%\label{demon}
%\end{figure}

The universality class of quantum criticality is determined by the critical exponents.
As we have seen in Fig.~\ref{phase-diagram}, the 1D attractive Hubbard model has a rich phase diagram.
At least one branch of the density of states shows sudden change when the driving parameters vary across the phase boundary in the phase diagram.
The singular behavior of thermodynamic properties is uniquely determined by the critical exponents,  which are independent of the microscopic details of the system.
Indeed, quantum criticality of quantum many-body systems  depends  solely  on the  dimensionality and the symmetry of the Hamiltonian.
Here we expand the above equations of state for the thermodynamic quantities in the limit $|\mu-\mu_c| \ll T$.
We  derive the scaling forms of the thermodynamics at quantum criticality and thus read off the critical exponents.

We find that the suddenly changed density of state usually results in a quantum phase transition,  so that the thermodynamical properties can be cast into  the forms of universal quantum scaling functions in the critical region.
For example,  for the phase transition from the fully-paired phase V to the FFLO-like state IV,
 thermodynamic quantities of  excess fermions  display   the singular parts in the scaling functions,
 whereas the thermodynamic properties of the bound pairs present  the  regular parts.
In contrast to the attractive SU(2) Fermi gas, the half-filling  phase in the attractive Hubbard model
contributes a constant regular part to the thermodynamic quantities due to its unique band-filling.

Our results for the scaling functions of particle density, magnetization, compressibility and susceptibility are summarized as follows:\\
$\bullet$ phase transition  (I-V),
\begin{align}
n_c=&-\sqrt{\frac{2|u|}{\pi}} \, T^{\frac{1}{2}} \, \textrm{Li}_{\frac{1}{2}} \left( -\exp\left( \frac{2\mu-2\mu_{c1}}{T} \right) \right),\notag\\
m\approx & \, 0 ,\notag\\
\kappa=& -2 \sqrt{\frac{2|u|}{\pi}} \, T^{-\frac{1}{2}} \, \textrm{Li}_{-\frac{1}{2}} \left( -\exp\left( \frac{2\mu-2\mu_{c1}}{T} \right) \right),\notag\\
\chi \approx & \, 0.
\end{align}
$\bullet$ phase transition (I-II),
\begin{align}
n_c=& -\frac{1}{2\sqrt{\pi}} T^{\frac{1}{2}} \textrm{Li}_{\frac{1}{2}} \left(-\exp \left( \frac{\mu-\mu_{c2}}{T} \right)  \right), \notag\\
m=&-\frac{1}{4\sqrt{\pi}} T^{\frac{1}{2}} \textrm{Li}_{\frac{1}{2}} \left(-\exp \left( \frac{\mu-\mu_{c2}}{T} \right)  \right), \notag\\
\kappa=&-\frac{1}{2\sqrt{\pi}} \, T^{-\frac{1}{2}} \textrm{Li}_{-\frac{1}{2}} \left( -\exp\left( \frac{\mu-\mu_{c2}}{T} \right) \right), \notag\\
\chi=& -\frac{1}{4\sqrt{\pi}} \, T^{-\frac{1}{2}} \textrm{Li}_{-\frac{1}{2}} \left( -\exp\left( \frac{\mu-\mu_{c2}}{T} \right) \right).
\end{align}
$\bullet$ phase transition (II-III),
\begin{align}
n_c=&\, 1+\frac{1}{2\sqrt{\pi}} \, T^{\frac{1}{2}} \, \textrm{Li}_{\frac{1}{2}} \left( - \exp\left( -\frac{\mu-\mu_{c3}}{T} \right) \right),\notag\\
m=&\frac{1}{2}+\frac{1}{4\sqrt{\pi}} \, T^{\frac{1}{2}} \, \textrm{Li}_{\frac{1}{2}} \left( - \exp\left( -\frac{\mu-\mu_{c3}}{T} \right) \right),\notag\\
\kappa=& -\frac{1}{2\sqrt{\pi}} \, T^{-\frac{1}{2}} \textrm{Li}_{-\frac{1}{2}} \left( - \exp\left( -\frac{\mu-\mu_{c3}}{T} \right) \right),\notag\\
\chi=& -\frac{1}{4\sqrt{\pi}} \, T^{-\frac{1}{2}} \textrm{Li}_{-\frac{1}{2}} \left( - \exp\left( -\frac{\mu-\mu_{c3}}{T} \right) \right).
\end{align}
$\bullet$ phase transition (II-IV),
\begin{align}
n_c=&n_{b4}+\lambda_1 T^{\frac{1}{2}}  \textrm{Li}_{\frac{1}{2}} \left( -\exp\left( \frac{2(\mu-\mu_{c4})}{T} \right) \right),\notag\\
m=&m_{b4}+\lambda_2 T^{\frac{1}{2}} \textrm{Li}_{\frac{1}{2}} \left( -\exp\left( \frac{2(\mu-\mu_{c4})}{T} \right) \right),\notag\\
\kappa=&\kappa_{b4}+\lambda_3 T^{-\frac{1}{2}}\textrm{Li}_{-\frac{1}{2}} \left(-\exp\left( \frac{2(\mu-\mu_{c4})}{T} \right) \right),\notag\\
\chi=&\chi_{b4}+\lambda_4 \, T^{-\frac{1}{2}}  \textrm{Li}_{-\frac{1}{2}} \left( -\exp\left(\frac{2(\mu-\mu_{c4})}{T} \right) \right).
\end{align}
$\bullet$ phase transition (V-IV),
\begin{align}
n_c=&n_{b5}+\lambda_5 \, T^{1/2} \textrm{Li}_{1/2} \left( -\exp \left( \frac{\mu-\mu_{c5}}{T} \right) \right),\notag\\
m=& \,-\frac{1}{4\sqrt{\pi}} \, T^{1/2} \, \textrm{Li}_{1/2} \left( -\exp \left( \frac{\mu-\mu_{c5}}{T} \right) \right),\notag\\
\kappa=& \kappa_{b5}+ \lambda_6 \, T^{-1/2} \, \textrm{Li}_{-1/2} \left( -\exp \left( \frac{\mu-\mu_{c5}}{T} \right)  \right),\notag\\
\chi=& \, -\frac{1}{4\sqrt{\pi}} \, T^{-1/2} \textrm{Li}_{-1/2} \left( -\exp\left( \frac{\mu-\mu_{c5}}{T} \right) \right).
\end{align}

\begin{figure}[!t]
\centering
\subfloat[]{\label{qc12-mag}\includegraphics[width = 0.75\columnwidth]{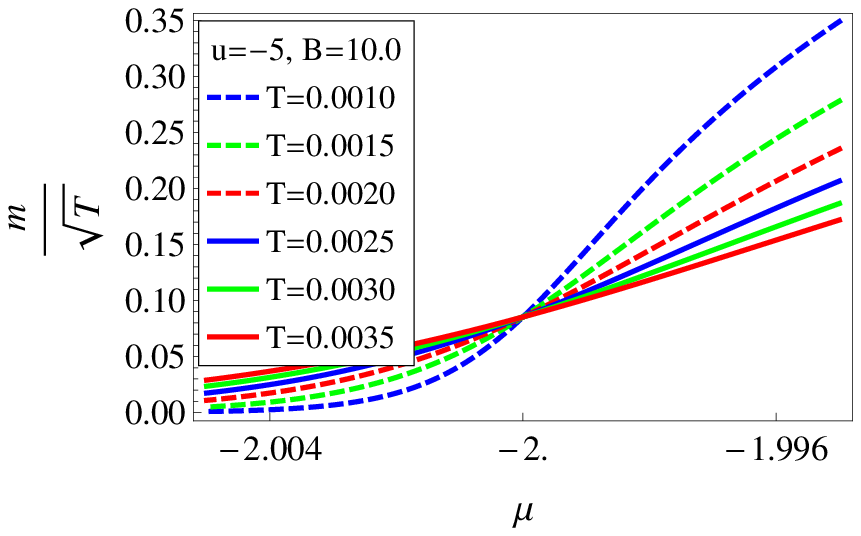}}\\
\subfloat[]{\label{qc23-sus}\includegraphics[width = 0.75\columnwidth]{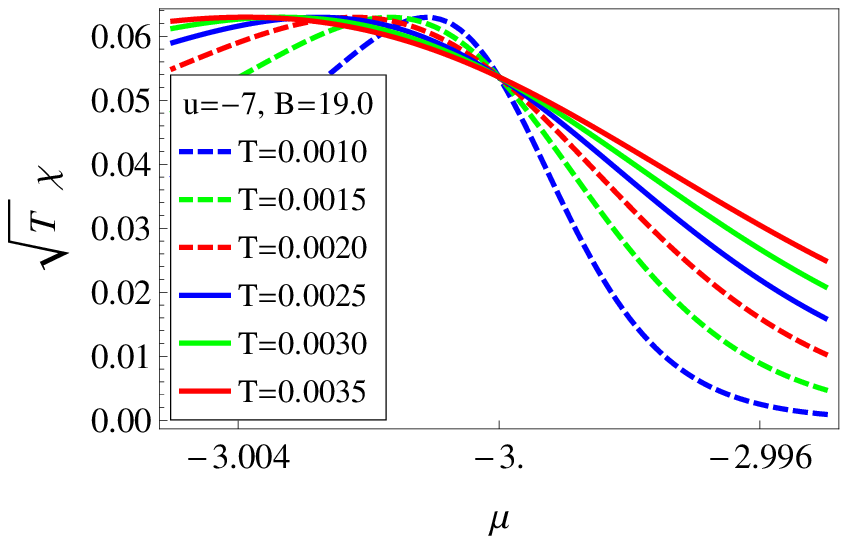}}\\
\subfloat[]{\label{qc24-den}\includegraphics[width = 0.75\columnwidth]{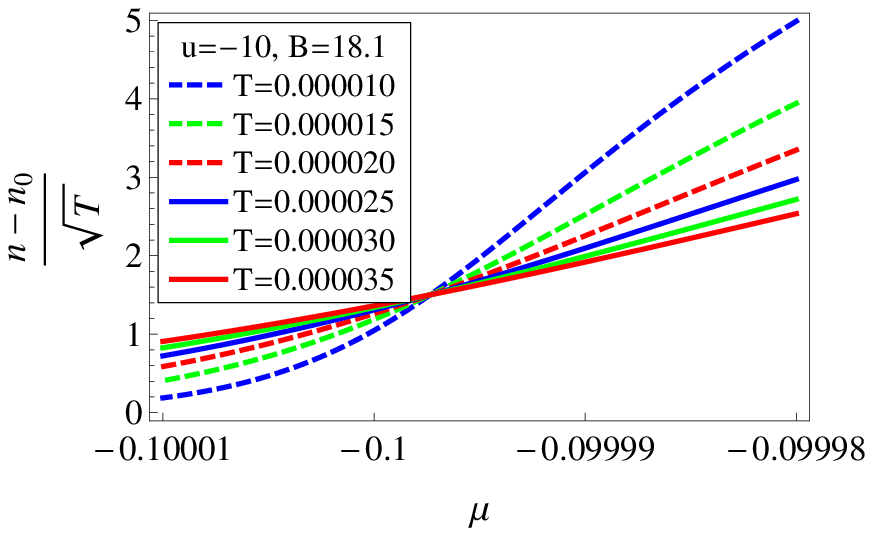}}\\
\subfloat[]{\label{qc15-comp}\includegraphics[width = 0.75\columnwidth]{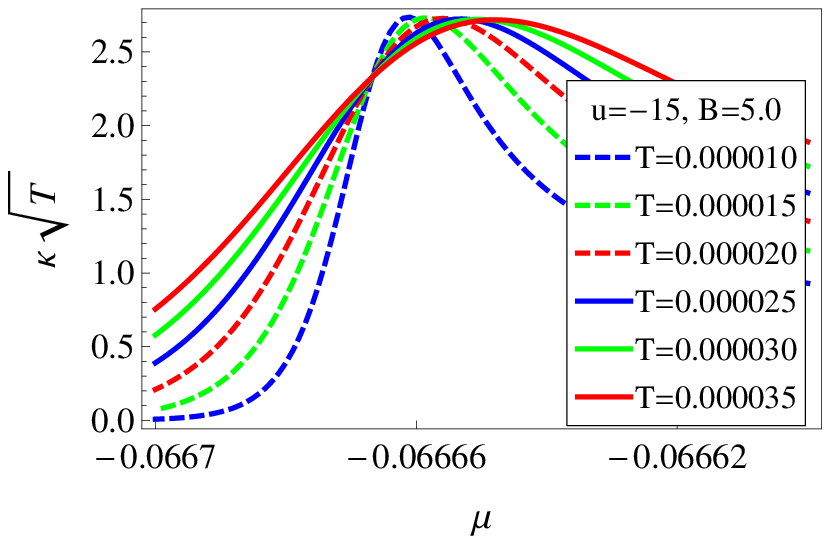}}
\caption{Scaling laws for thermodynamic quantities vs. chemical potential at different temperatures.
The intersection points  in (a), (b), (c) and (d) give the critical points for phase transitions (I-II), (II-III), (II-IV) and (I-V),  respectively. }
\label{fig:scalings}
\end{figure}

In the above scaling forms some constants are given in Appendix \ref{Scaling-Constants}.
These  scaling forms can be cast into the form of well known universal scaling laws.
For example, the universal scaling laws for  the density and compressibility read \cite{Fisher1989,Sachdev,Hazzard2011,Ho2010}
\begin{align}
n(\mu,B,T)=&\, n_0(\mu,B,T)+T^{d/z+1-(1/\nu z)} \mathcal{G}\left( \frac{\mu-\mu_c}{T^{1/\nu z}} \right), \notag\\
\kappa(\mu,B,T)=&\, \kappa_0(\mu,B,T)+T^{d/z+1-(2/\nu z)} \mathcal{F}\left( \frac{\mu-\mu_c}{T^{1/\nu z}} \right),
\end{align}
where $n_0$ and $\kappa_0$ are the regular parts,  i.e., the background values before the phase transition.
Meanwhile $\mathcal{G}(x)=\textrm{Li}_{\frac{1}{2}}(x)$, $\mathcal{F}(x)=\textrm{Li}_{-\frac{1}{2}}(x)$ give the scaling functions in the singular parts.
From the above scaling forms, we  read off the dynamical exponent $z=2$ and correlation critical exponent $\nu=1/2$.
This scaling theory is valid for all  phase transitions across the phase boundaries in the phase diagram \ref{phase-diagram}.
Such universal scaling laws are demonstrated in Fig.~\ref{fig:scalings} for various phase transitions.

The above scaling forms are observed to give the same critical exponents which characterize the universality class of free-fermion criticality.
An intuitive explanation for this result is that the phase transitions occurred in the 1D Hubbard model  have a common feature:  at least one branch of Fermi sea vanishing, namely $\varepsilon^{u,b}(0) =0$.
This naturally leads to a change in dispersion, i.e., a linear dispersion vanishes while a quadratic dispersion is created when the phase transition occurs.
This change in dispersion underlies a universality class of quantum criticality, see also the recent studies of the 1D interacting Bose gas \cite{Yang:2017}
and the 1D Heisenberg spin chain \cite{He:2017}.

Moreover, the phase V in the phase diagram Fig.~\ref{phase-diagram} shows a gapped phase (fully paired phase),
where the susceptibility reveals a particular exponential decay at low temperatures.
Using the equation of state with the pressures (\ref{eos-pu}), (\ref{eos-pb}), and (\ref{eos-d1})-(\ref{eos-phi}),
we further  show  that the susceptibility decays exponentially with the energy gap induced by the ferromagnetic ordering, namely
\begin{equation}
\chi \approx \frac{T^{-1/2}}{4\sqrt{\pi}}e^{-\Delta/T},
\end{equation}
where the energy gap is given by $\Delta=\varepsilon^u(0)=-2-\mu-2u-B+p^b$ with  $p^b=\frac{4(2\pi-q^3/3)}{3|u|\pi}\left( 1+\frac{2|u|\pi\mu}{2\pi-q^3/3}\right)^\frac{3}{2}$.
This result can also be obtained by applying Sommerfeld expansion in  \cref{effective-pb}.
We  approximately obtain   the susceptibility
$\chi \approx \frac{1}{2} \gamma' \approx -\frac{T^{-1/2}}{2\sqrt{\pi}} \textrm{Li}_{-\frac{1}{2}}\left(-\mathrm{e}^{-\Delta/T}\right)$ from \cref{sus}.
In the next section, we further demonstrate the macroscopic nature of the susceptibility in the FFLO phase.

\section{Free Fluids and Additivity Rules}\label{sec-add}

Fermi liquid theory is believed to break down in 1D strongly correlated systems due to the absence of well defined quasi-particles \cite{Giamarchi:2004}.
Consequently the TLL theory is generally believed to describe the collective low-lying excitations in 1D many-body systems.
Despite such a big difference in the microscopic origins of the two low-energy theories,
both the Fermi liquid and the TLL share a common feature -- a small  distortion of the Fermi surface or Fermi points results in
the universal low-energy physics of many-body systems.
From the results of the last section, we observed that at very low temperatures the low-energy physics of the FFLO-like  state
is governed by the universality class of a two-component TLL.
However, in view of  the macroscopic properties of the 1D attractive Hubbard model,
we argue that such a universality class of two-component TLL reveals an important nature of  free fluids.
In order to show this elegant nature, we will introduce two  effective chemical potentials for the excess fermions and bound pairs on a 1D lattice.
Then we will show  that the thermodynamic properties in the FFLO-like phase   behave like  two independent free fluids.
In particular, we find simple  additivity rules for the compressibility and susceptibility which represent a universal characteristic of quantum liquids at the renormalization fixed point.

Prior to a discussion of the free fluids, we first make an approximation for the zero temperature TBA equations \cref{0tba1,0tba2} in the low density regime,
\begin{align}
\varepsilon^1(k)=&\,k^2-\mu_1 -a_1 \star \varepsilon^2(k), \label{tba_simp_lowT_lowN_final_1} \\
\varepsilon^2(\Lambda)=&\,\alpha_1 \Lambda^2-\alpha_1 \mu_2 - a_1 \star \varepsilon^1(\Lambda) - a_2 \star \varepsilon^2(\Lambda),\label{tba_simp_lowT_lowN_final_3}
\end{align}
where $a_m\star \varepsilon^n({x})=\int_{-y_c}^{y_c} \textmd{d}y \, a_m(x-y)\varepsilon^n(y)$ with $y_c$ being the Fermi point of $\varepsilon^n(y)$, i.e., $\varepsilon^n(y_c)=0$.
In the above equations, we introduced the effective chemical potentials for excess fermions and bound pairs as
\begin{align}
\mu_1=&\mu+B+2u+2,\label{effective-mu1}\\
\mu_2=&\frac{2}{\alpha_1}\left(\mu+2\sqrt{u^2+1}-2|u| \right).\label{effective-mu2}
\end{align}
The effective chemical potential of the bound pairs  reveals a deep physical insight into the crossover from Bose-Einstein condensate (BEC) to Bardeen-Cooper-Schrieffer (BCS) superconductor.
Later we shall see these effective chemical potentials reveal an important free quantum liquid nature.
We will show  that for the balanced case the effective chemical potential $\mu_2$ varies from the kinetic energy of bound pairs to the free Fermi energy when the interaction changes from negative infinity to zero.
This reveals a 1D analogue of the BEC-BCS crossover.
This form of  the TBA equations is useful to access the ground state properties, such as sound velocities, stiffness and effective chemical potentials.
By virtue of Eqs. (\ref{tba_simp_lowT_lowN_final_1}) and  (\ref{tba_simp_lowT_lowN_final_3}) we rewrite the free energy per site (\ref{free3}) as
\begin{align}\label{free_energy_simp_lowT_lowN_final}
f=u+\int_{-k_c}^{k_c} \frac{\textmd{d}k}{2\pi} \varepsilon^1 (k) + \int_{-\Lambda_c}^{\Lambda_c} \frac{\textmd{d}\Lambda}{2\pi} \beta_1 \varepsilon^2(\Lambda),
\end{align}
where $\varepsilon^u=\varepsilon^1$ and $\varepsilon_1^\prime=\varepsilon^2$.

We now proceed to  calculate  the TLL parameters of the model and compare them with those of the 1D  attractive SU(2) Fermi gas \cite{YCY2016}.
The basic idea is to express the effective chemical potentials in terms of the Fermi points by employing iteration
of \cref{tba_simp_lowT_lowN_final_1,tba_simp_lowT_lowN_final_3}.
Using  the fact  that the two  dressed energies vanish at their corresponding Fermi points, we  express those Fermi points in terms of  the densities of excess fermions and bound pairs.
We shall  see that this process leads to a separation of two free fluids in the ground state energy per site.

In order to simplify the lengthy iterations, we firstly rescale the TBA equations (\ref{tba_simp_lowT_lowN_final_1}) and (\ref{tba_simp_lowT_lowN_final_3}) by defining $\tilde{\varepsilon}^n={\varepsilon^n}/u^2$, $\tilde{\mu}_n=\mu_n/u^2$, $\tilde{y}_c=y_c/|u|$, and $\tilde{a}_n(x)=\frac{n}{\pi} \frac{1}{n^2+x^2}$. Then we introduce a vector presentation of the rescaled TBA equations.
In view of the properties of even functions, we utilize the base $\left\{k^{2n}\right\}$ and $\left\{\Lambda^{2n}\right\}$ ($n=0,1,2,\ldots$)
to expand these scalar equations, thus we  have,  respectively
\begin{align}
\vec{\varepsilon}^{\,1}=& \, \vec{V}^1-\mathbf{A}^1(\tilde{\Lambda}_c)\,\vec{\varepsilon}^{\,2},\label{vec-tba1}\\
\vec{\varepsilon}^{\,2}=& \, \vec{V}^2-\mathbf{A}^1(\tilde{k}_c)\,\vec{\varepsilon}^{\,1}-\mathbf{A}^2(\tilde{\Lambda}_c)\,\vec{\varepsilon}^{\,2}.\label{vec-tba2}
\end{align}
The vectors  $\vec{V}^1=\left[-\tilde{\mu}_1,1,0,\ldots \right]^t$ and $\vec{V}^2=\left[-\alpha_1\tilde{\mu}_2,\alpha_1,0,\ldots \right]^t$
are the driving terms and the  superscript $t$ represents transpose operation.
The matrix $\mathbf{A}^n(\tilde{y}_c) \vec{\varepsilon}$ corresponds to the integral $\int_{-\tilde{y}_c}^{\tilde{y}_c} \textmd{d}y \, \tilde{a}_n(x-y) \tilde{\varepsilon}(y)$.

Furthermore, as we only retain the first few leading terms, $\vec{\varepsilon}^{\,n}$ and $\mathbf{A}^n(\tilde{y}_c)$
can be expanded as sums of a few leading orders with  respect to $y_c$, i.e.,
$\vec{\varepsilon}^{\,n}=\vec{\varepsilon}_{(0)}^{\,n}+\vec{\varepsilon}_{(1)}^{\,n}+\vec{\varepsilon}_{(2)}^{\,n}+\ldots$ and
$\mathbf{A}^n(\tilde{y}_c)=\mathbf{A}_{(1)}^n(\tilde{y}_c)+\mathbf{A}_{(3)}^n(\tilde{y}_c)+\mathbf{A}_{(5)}^n(\tilde{y}_c)+\ldots$.
More details of the latter expansion are presented  in \cref{app-iteration}.
Substituting  these expansions into the TBA equations of vectorial form \cref{vec-tba1,vec-tba2},
and sorting  terms order by order,  leads to   the set of  equations
%
%\begin{widetext}
\begin{align}
\vec{\varepsilon}_{(0)}^{\,1}=& \, \vec{V}^1, \qquad \vec{\varepsilon}_{(0)}^{\,2}=\, \vec{V}^2, \notag\\
\vec{\varepsilon}_{(1)}^{\,1}=&-\mathbf{A}^1_{(1)}(\tilde{\Lambda}_c) \vec{\varepsilon}_{(0)}^{\,2}, \notag\\
\vec{\varepsilon}_{(1)}^{\,2}=&-\mathbf{A}^1_{(1)}(\tilde{k}_c) \vec{\varepsilon}_{(0)}^{\,1} - \mathbf{A}^2_{(1)}(\tilde{\Lambda}_c) \vec{\varepsilon}_{(0)}^{\,2}, \notag\\
\vec{\varepsilon}_{(2)}^{\,1}=&-\mathbf{A}^1_{(1)}(\tilde{\Lambda}_c) \vec{\varepsilon}_{(1)}^{\,2}, \notag\\
\vec{\varepsilon}_{(2)}^{\,2}=&-\mathbf{A}^1_{(1)}(\tilde{k}_c) \vec{\varepsilon}_{(1)}^{\,1} - \mathbf{A}^2_{(1)}(\tilde{\Lambda}_c) \vec{\varepsilon}_{(1)}^{\,2}, \notag\\
\vec{\varepsilon}_{(3)}^{\,1}=&-\mathbf{A}^1_{(1)}(\tilde{\Lambda}_c) \vec{\varepsilon}_{(2)}^{\,2} - \mathbf{A}^1_{(3)}(\tilde{\Lambda}_c) \vec{\varepsilon}_{(0)}^{\,2},  \notag\\
\vec{\varepsilon}_{(3)}^{\,2}=&-\mathbf{A}^1_{(1)}(\tilde{k}_c) \vec{\varepsilon}_{(2)}^{\,1} - \mathbf{A}^2_{(1)}(\tilde{\Lambda}_c) \vec{\varepsilon}_{(2)}^{\,2} - \mathbf{A}^1_{(3)}(\tilde{k}_c) \vec{\varepsilon}_{(0)}^{\,1} \notag\\
&- \mathbf{A}^2_{(3)}(\tilde{\Lambda}_c) \vec{\varepsilon}_{(0)}^{\,2}.
\end{align}
%\end{widetext}
%
It is easy to solve the above vectorial forms $\vec{\varepsilon}_{(r)}^{\,1}$ and $\vec{\varepsilon}_{(r)}^{\,2}$ with $r=1,2,3$.
We then substitute these results into the scalar expression of the rescaled TBA equations.
Together with  $\tilde{\varepsilon}^n(\tilde{y}_c)=0$ and the expansion $\tilde{\mu}_n=\tilde{\mu}_n^{(2)}+\tilde{\mu}_n^{(3)}+\tilde{\mu}_n^{(4)}+\cdots$,
we then obtain a set of recurrence equations for $\tilde{\mu}_n^{(2)}$, $\tilde{\mu}_n^{(3)}$ and $\tilde{\mu}_n^{(4)}$.
Here we observe that the expansions for chemical potentials begin from $n=2$ due to the fact that $\tilde{\varepsilon}^1(\tilde{k}_c)=-\tilde{\mu}_1+\tilde{k}_c^2+o\left( \tilde{k}_c^3\right)=0$, i.e.,  $\tilde{\mu}_1=\tilde{k}_c^2+o\left( \tilde{k}_c^3\right)$. Similarly for $\tilde{\mu}_2$.
We solve these equations and then express the solution as the vectorial equation
\begin{align}
\left[
\begin{array}{c}
 \tilde{\mu}_1 \\
 \alpha_1 \tilde{\mu}_2 \\
\end{array}
\right]
=\left(\mathbf{I}+\frac{2}{3}\mathbf{T}\right)
\left[
\begin{array}{c}
 \tilde{k}_c^2 \\
\alpha_1 \tilde{\Lambda}_c^2 \\
\end{array}
\right],
\label{solution_mu_matrix_hubbard}
\end{align}
where the matrix $\mathbf{T}$ is given by
\begin{align}
\mathbf{T} =\frac{1}{\pi}
\left[
\begin{array}{cc}
 0 & 2\tilde{\Lambda}_c \\
 2\tilde{k}_c & \tilde{\Lambda}_c \\
\end{array}
\right].
\label{definition_T_matrix}
\end{align}
Finally, the free energy per site \cref{free_energy_simp_lowT_lowN_final}  is expressed in terms of Fermi points $k_c$ and $\Lambda_c$, with result
\begin{align}
f=-\frac{2}{3\pi} \left(k_c^3+\alpha_1\beta_1 \Lambda_c^3\right)+u.
\label{free_energy_hubbard_result_2}
\end{align}

We now proceed  to obtain the particle densities  in terms of  the Fermi points.
To this end,  we turn to the total particle density $n_c$ and magnetization $\bar{m}$ per site based on \cref{free_energy_simp_lowT_lowN_final},
\begin{align}
n_c=&-\frac{\partial f}{\partial \mu}= -\int_{-k_c}^{k_c} \frac{\textmd{d}k}{2\pi} \frac{\partial \varepsilon^1}{\partial \mu} - \beta_1 \int_{-\Lambda_c}^{\Lambda_c}
\frac{\textmd{d}\Lambda}{2\pi}\frac{\partial \varepsilon^2}{\partial \mu}, \notag\\
\bar{m}=&-\frac{\partial f}{\partial B}= -\int_{-k_c}^{k_c} \frac{\textmd{d}k}{2\pi} \frac{\partial \varepsilon^1}{\partial B} - \beta_1 \int_{-\Lambda_c}^{\Lambda_c}
\frac{\textmd{d}\Lambda}{2\pi}\frac{\partial \varepsilon^2}{\partial B}. \notag
\end{align}
In order to get  closed forms for these two properties,
we  first take partial derivatives of \cref{tba_simp_lowT_lowN_final_1,tba_simp_lowT_lowN_final_3} with respect to $\mu$ and $B$, respectively.
Then we   rewrite these integral equations in terms of the vectorial forms similar to  \cref{vec-tba1,vec-tba2}.
Finally, by lengthy  iteration and after some manipulations, we obtain
\begin{align}\label{density-by-Fermi-point}
\left[
  \begin{array}{c}
    \tilde{n}_1 \\
    \tilde{n}_2 \\
  \end{array}
\right]
=\frac{1}{\pi}\left( \mathbf{I}-\mathbf{T}+\mathbf{T^2}  \right)^t
\left[
  \begin{array}{c}
    \tilde{k}_c \\
    \beta_1 \tilde{\Lambda}_c \\
  \end{array}
\right],
\end{align}
where $\tilde{n}_r=n_r/|u|$ ($r=1,2$) with $n_1=\bar{m}$ and $n_2=\left(n_c-n_1\right)/2$ being respectively the densities for the excess fermions and bound pairs.
Here the redefined $\bar{m}=2m$ is introduced according to the original TBA equations.
 An inverse of \cref{density-by-Fermi-point} gives the cut-off momenta in terms of the densities $n_{1,2}$
\begin{eqnarray}
k_c& \approx& \pi n_1 \sum_{n=0}^3 \left( \frac{2n_2}{|u|} \right)^n, \nonumber\\
\Lambda_c & \approx &\frac{\pi n_2}{\beta_1} \sum_{n=0}^3 \left[ \frac{2n_1+n_2}{\beta_1|u|} \right]^n.\label{Fermi-point-dens}
\end{eqnarray}

For the next step, substituting Eq. (\ref{Fermi-point-dens}) into (\ref{free_energy_hubbard_result_2}),
leads to separating the ground state energy per site into the energies of excess fermions and bound pairs, with result
\begin{align}
e=e_1+e_2+e_{b}.
\end{align}
Here $e_b$ is the binding energy and the subscripts $1$ and  $2$ denote the  excess fermions and  bound pairs, respectively.
The terms are given explicitly by
\begin{align}
e_1=&\frac{\pi^2}{3}\, n_1^3 \left[1+2\left(\frac{2n_2}{|u|}\right)+3\left(\frac{2n_2}{|u|}\right)^2\right], \label{result_E1_hubbard}\\
e_2=&\frac{\pi^2}{3}\, \frac{\alpha_1 n_2^3}{\beta_1^2}\left[1+2\left(\frac{2n_1+n_2}{\beta_1|u|}\right)+3\left(\frac{2n_1+n_2}{\beta_1|u|}\right)^2\right], \label{result_E2_hubbard} \\
e_{b}=&-\left(2u+2\right)n_1-4\left(u+\sqrt{u^2+1}\right)n_2. \label{result_Ebond_hubbard}
\end{align}
As usual,  we define a dimensionless interaction strength  $\gamma_s={2|u|}/{n_s}$ ($s=1,2$) \cite{YCY2016}. Using the relation
\begin{equation}
K_s={\pi}/{\sqrt{3e(\gamma_s)-2\gamma_s \frac{\textmd{d}e(\gamma_s)}{\textmd{d}\gamma_s}+\frac{1}{2}\gamma_s^2\frac{\textmd{d}^2 e(\gamma_s)}{\textmd{d}\gamma_s^2}}},
\end{equation}
the Luttinger parameters for the excess fermions and bound pairs can be directly worked out to be
\begin{align}\label{TLL-parameter}
K_1=1,\,\, \, K_2=2\sqrt{2}\,\frac{\beta_1}{\sqrt{\alpha_1}} \left[1-\frac{2}{\beta_1\,\gamma_2}+\frac{1}{(\beta_1\,\gamma_2)^2} \right].
\end{align}

We note that the Luttinger parameter $K_2$ in the fully paired phase depends explicitly on the lattice parameters $\alpha_1$ and $\beta_1$.
This behavior is different from the constant value $K_2=4$ for the bound pairs phase of the strongly attractive SU(2) Fermi gas \cite{YCY2016}.
In the limits  $u\rightarrow 0$ and $n_s/|u|$  small,  the lattice parameters $\alpha_1\rightarrow 2$, $\beta_1\rightarrow 2$.
Thus  we have  $K_2=4$ which is the same as for the SU(2) Fermi gas.
The two limits $u \rightarrow 0$ and $n_s/|u|\ll 1$  represent the lattice-gas mapping  between 1D attractive Hubbard model and SU(2) Fermi gas \cite{Krivnov1975}.

Beside this framework of the TLL theory, we also find that for  the  low density case,
the chemical potentials for the unpaired fermions and pairs are given explicitly by
\begin{eqnarray}
\mu_1&=& \pi n_1^2A_1^2+\frac{4\pi^2\alpha_1}{3\beta_1^3|u|}n_2^3A_2^3, \label{chemical-1}\\
\mu_2&=&\pi^2\frac{n_2^2}{\beta_1^2}A_2^2+\frac{4\pi^2}{3\alpha_1|u|}n_1^3A_1^3+\frac{2\pi^2}{3\beta_1^3|u|}n_2^3A_2^3 \label{chemical-2},
\end{eqnarray}
where    $A_1=1+\frac{2n_2}{|u|}+\left(\frac{2n_2}{|u|}\right)^2$ and $A_2=1+\frac{2n_1+n_2}{\beta_1 |u|}+\left(\frac{2n_1+n_2}{\beta_1 |u|}\right)^2$,
which  indicate  interacting effects among pairs and unpaired fermions.
We observe that the chemical potential $\mu_2$ tends to  the kinetic energy of bound pairs  in the BEC limit $|u| \to \infty$.
Whereas in the weak coupling limit, $|u| \to 0$, $\mu_2$ tends to  the Fermi energy of the free fermions on a 1D lattice.
%
%Specifically, $A_1$ and $A_2$ represent  the statistical parameters for ideal anyonic particles of single atoms and pairs, respectively.
%
The effective chemical potentials (\ref{chemical-1}) and (\ref{chemical-2}) reveal  that the thermodynamic quantities could be separable,
i.e., the total is equal to a sum of the effective thermodynamic quantities of  two individual constituents.

\begin{figure}[ht]
\centering
\includegraphics[width=0.35\textwidth]{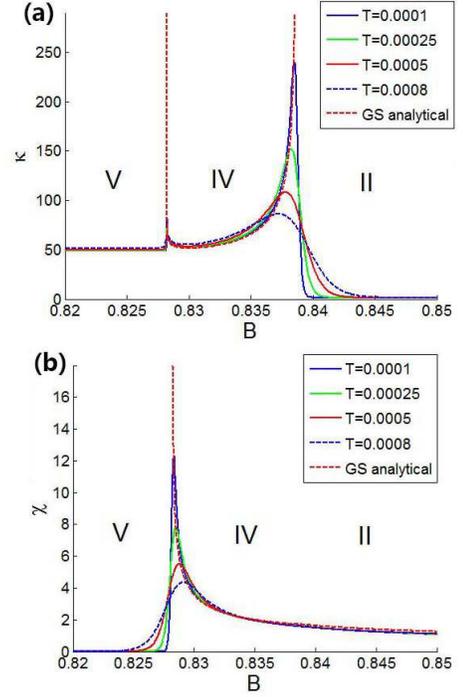}
\caption{  Additivity rules: (a) Compressibility $\kappa$ and (b) spin susceptibility $\chi$  vs. magnetic field $B$  for the attractive Hubbard  model with   $u=-1$ and $\mu=-0.8282$.
The red dashed lines show the result obtained from the  additivity rules (\ref{add-kappa})  and (\ref{add-sus}).
At low temperatures, all  compressibility and susceptibility curves collapse into the zero temperature ones obeying the  additivity rules.
In the vicinity of the critical points such  free fluids nature beaks down.
 }
\label{fig:additivity}
\end{figure}

Here we further derive the additivity rules for the compressibility and susceptibility.
For the compressibility, using the standard thermodynamic relation $\kappa=\left(\frac{\partial n_c}{\partial \mu}\right)_B$,
the derivatives of the density and effective chemical potentials for fixed magnetic field could be further expressed as
$\textmd{d}n_c=\textmd{d}n_1+2\,\textmd{d}n_2$ and $\textmd{d}\mu_1=\frac{\alpha_1}{2}\textmd{d}\mu_2=\textmd{d}\mu$, respectively.
Inserting these relations into the definition of compressibility, $\kappa=\frac{\partial n_c}{\partial \mu}\Big{|}_B=\frac{\texttt{d}n_1+2\texttt{d}n_2}{\texttt{d}\mu}$,
we thus obtain
\begin{align}
\kappa=\kappa_1+\frac{2}{\alpha_1}\kappa_2.\label{add-kappa}
\end{align}
Here the effective compressibilities of excess fermions and bound pairs are defined as
$\kappa_1=\left(\frac{\partial n_1}{\partial \mu_1}\right)_B$ and $\kappa_2=2\left(\frac{\partial n_2}{\partial \mu_2}\right)_B$.
Details are given in see in Appendix  \ref{app-iteration}.
The additivity rule (\ref{add-kappa}) for the compressibility can be confirmed numerically, as shown  in Fig.~\ref{fig:additivity}(a).

For the susceptibility in the canonical ensemble, defined as $\bar{\chi}=\left(\frac{\partial \bar{m}}{\partial B}\right)_{n_c}$,
it is straightforward to see $\textmd{d}n_c=\textmd{d}n_1+2\textmd{d}n_2=0$ and $\textmd{d}B=\textmd{d}\mu_1-\frac{\alpha_1}{2}\textmd{d}\mu_2$,
and thus the additivity rule
\begin{align}
\frac{1}{\bar{\chi}}=\frac{1}{\bar{\chi}_1}+\frac{\alpha_1}{2}\frac{1}{\bar{\chi}_2}.\label{add-sus}
\end{align}
Here $\bar{\chi}_1=\left(\frac{\partial n_1}{\partial \mu_1}\right)_{n_c}$ and $\bar{\chi}_2=2\left(\frac{\partial n_2}{\partial \mu_2}\right)_{n_c}$
are the effective susceptibilities for excess fermions and bound pairs, respectively.
These explicit expressions for the effective thermodynamic quantities can be found in Appendix \ref{app-iteration}.
The additivity rule (\ref{add-sus}) for the susceptibility can  also be confirmed  numerically, as shown in Fig.~\ref{fig:additivity}(b).
Similar to the observation concerning TLL parameters,
the additivity rules for the 1D attractive  Hubbard model also reduce to those for the SU(2) Fermi gas through the lattice-gas mapping.
In Appendix  \ref{app-iteration} we calculate the individual compressibility and susceptibility explicitly.

\begin{figure}[!t]
\centering
\includegraphics[width=0.45\textwidth]{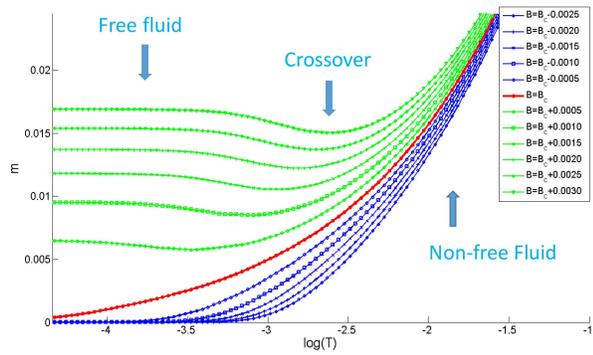}
\caption{
Numerical results for the magnetization vs. logarithm of the temperature
for  different magnetic fields. Here we have set a  fixed chemical potential $\mu=-0.14$ and interaction strength  $u=-7$.
For magnetic field $B>B_c=12.11065$ (phase IV), three regions are clearly displayed:
The free fluids  region at  low temperatures, non-Fermi liquid region at  higher temperatures, and a crossover in between.
For magnetic field $B<B_c$ (phase V), the magnetization displays the gapped nature of a non-Fermi liquid phase.   }
\label{fig:FL}
\end{figure}

The simple additivity nature of the thermodynamics at low temperatures
characterizes the universal low energy physics of the FFLO-like state of the 1D attractive Hubbard model.
In this sense, the additivity rules reflect a universal nature of the multicomponent  TLL in 1D.
The simple additivity rule thus reveals the significant two free fluid nature of the FFLO phase,
as predicted in expansion dynamics of the FFLO state in 1D \cite{Kajala:2011}.
The  macroscopic magnetic properties in the FFLO-like phase show the properties of the ordinary higher-dimensional Fermi liquid,
see Fig.~\ref{fig:FL}.
This figure shows that in  the free fluids  region the magnetization is nearly temperature independent.
In the non-Fermi liquid region thermal fluctuations gradually overwhelm quantum fluctuations.
 Thus the magnetization has a uniform temperature dependence for different magnetic fields, indicating paramagnetism.
The non-Fermi liquid crossover region reveals a scaling invariance, which was studied in Section IV.
Such Fermi liquid-like features have been found in the spin compound Cu(C${}_4$H${}_4$N${}_2$)(NO${}_3$)${}_2$ \cite{Kono:2015} and the heavy fermion material YbNi${}_4$P${}_2$ \cite{Krellner:2016}.
The study of Fermi and non-Fermi liquids in 1D has received significant recent interest \cite{Shaginyan:2016,Rozhkov:2014,Lebed:2015}.

Using the explicit expressions for the compressibility (\ref{add-kappa}) and susceptibility (\ref{add-sus}),
we may calculate the Wilson ratio, which is a dimensionless ratio defined as the susceptibility or compressibility over the specific heat divided by the temperature.
The Wilson ratio is the ratio describing quantum fluctuations and energy thermal fluctuations.
Both the Fermi liquid and TLL give a constant Wilson ratio \cite{YCY2016}, i.e., two types of fluctuations are on equal-footing in temperature scaling.
However, near a critical point, the dimensionless Wilson ratios exhibits a sudden enhancement indicating a sudden change in the density of state.
Therefore the Wilson ratios serves as a powerful tool for distinguishing the phases of a quantum liquid  and for determining the finite temperature phase diagram as well.

\begin{figure}[!t]
\centering
\includegraphics[width = 0.4\textwidth]{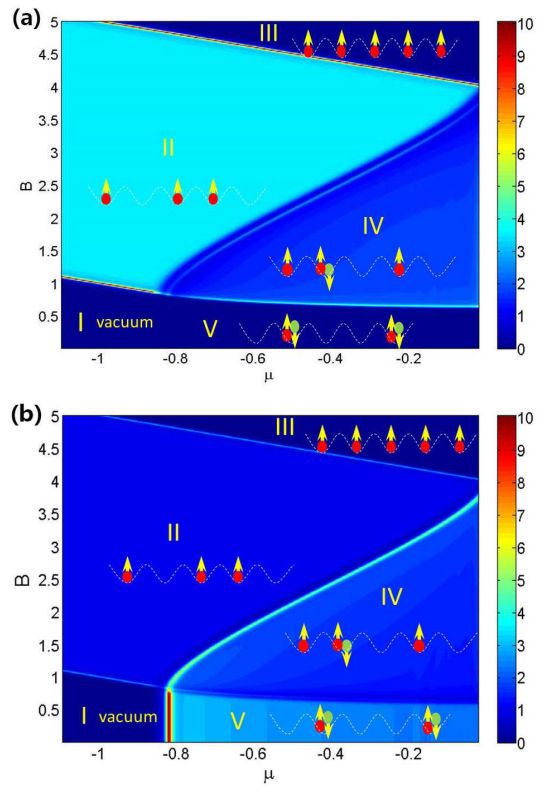}
\caption{
Finite temperature phase diagrams obtained from contour plots of the Wilson ratios.
The plots in (a) and (b) are determined from the susceptibility Wilson ratio $R_\mathrm{W}^\chi$
and from the compressibility Wilson ratio $R_\mathrm{W}^\kappa$, respectively.
Here $u=-1$ and $T=0.001$. The red balls and green balls represent up spin and down spin respectively.
Both diagrams agree well with the zero temperature phase diagram \Cref{phase-diagram},
despite the fact that the plot in (a) cannot distinguish phase I and V due to the vanishing susceptibility in these two phases.}
\label{wrcomb}
\end{figure}

The compressibility Wilson ratio $R_\mathrm{W}$ is determined by
\begin{eqnarray}
R_\mathrm{W}^\kappa&=&\frac{\pi^2 k_B^2}{3}\,\frac{\kappa}{C_v/T}\nonumber\\
&=&\pi \left(\kappa_1+\frac{2}{\alpha_1}\kappa_2\right)\bigg/\left( \frac{1}{v_1}+\frac{1}{v_2} \right),
\end{eqnarray}
where we have used \cref{low-tem-correction} to calculate the specific heat and set the Boltzmann constant to $k_B=1$.
This Wilson ratio vanishes in both phases I (vacuum) and III (half-filling phase).
In the limit $n_c/|u| \rightarrow 0$ the compressibility Wilson ratio for phases II and IV are respectively,
$R_\mathrm{W}^\kappa=1$ and $R_\mathrm{W}^\kappa=2\sqrt{2}\beta_1/\sqrt{\alpha_1}$.
These results turn out to be the same as for the  strongly attractive SU(2) Fermi gas \cite{YCY2016}
when  the limit $u \rightarrow 0$ is applied.
On the other hand, the susceptibility Wilson ratio is defined by
$R_\mathrm{W}^\chi=\frac{4}{3}\left(\frac{\pi k_B}{\mu_B\, g_L}\right)^2 \frac{\chi}{C_v/T}$
with Bohr magneton $\mu_B$ and Lande factor $g_L$.
Fig.~\ref{wrcomb} shows a contour plot of each type of Wilson ratio which demonstrates the macroscopic feature of the Fermi liquid nature.
This figure also presents the low-temperature phase diagram in the $B-\mu$ plane.

\section{Conclusion}
In summary we have presented a framework to determine the nature of quantum criticality and quantum liquids in the 1D attractive Hubbard model.
We have obtained the universal thermodynamics of the  model by solving  the TBA equations.
In particular, we have analytically derived the equation of state at low temperatures,
from which we have obtained effective chemical potentials of excess fermions and bound pairs, along with
the density, compressibility, susceptibility and specific heat in terms of the chemical potential $\mu$, magnetic field $B$,  temperature $T$ and interaction strength constant.
At quantum criticality  the  scaling forms of these thermal and magnetic properties have been obtained.
The dynamical exponent $z=2$ and correlation critical exponent $\nu=1/2$, indicating the universality class of criticality of free fermion theory.

Our results provide strong evidence for the existence of two free fluids of bound pairs and of unpaired fermions,
which were noticed in the expansion dynamics of the FFLO state in 1D \cite{Kajala:2011}.
Regarding the nature of the two fluids in the attractive Hubbard model,  we have  shown  that   in the low-density regime the interaction
effect resulting from the paired and unpaired fermions   can be absorbed into  effective chemical potentials
of two non-interacting ideal gases.
Consequently, the  additivity rules in the compressibility and susceptibility of the 1D attractive Hubbard model
hold as long as  the dimensionless Wilson ratio remains a constant.
This behavior significantly reflects the free fluids  nature in  thermodynamic properties of  the model.
In this phase, the FFLO pair correlation function
\begin{eqnarray}
G_p(x,t)&=&\langle \Psi_\uparrow^\dag(x,t) \Psi_\downarrow^\dag(x,t) \Psi_\uparrow(0,0) \Psi_\downarrow(0,0) \rangle \nonumber\\
&\approx& A_{p,1} \frac{\cos \left( \pi(n_\uparrow-n_\downarrow)x \right)}{|x+{\mathrm i}\, v_1 \, t|^{2\theta_1}\,|x+{\mathrm i}\, v_2 \, t|^{2\theta_2}}\nonumber\\
&&+ A_{p,2} \frac{\cos \left( \pi(n_\uparrow-3n_\downarrow)x \right)}{|x+{\mathrm i}\, v_1 \, t|^{2\theta_3}\,|x+{\mathrm i}\, v_2 \, t|^{2\theta_4}},
\end{eqnarray}
shows a typical spatial oscillation which  is a characteristic  of the FFLO state.
In the above equation, the exponents $\theta_1 \approx 1/2$, $\theta_2 \approx 1/2+\frac{n_2}{|u|\beta_1} $,  $\theta_3 \approx \frac{1}{2}-\frac{4\, n_2}{|u|\beta_1}$ and $\theta_4 \approx \frac{5}{2}-\frac{4\,n_1}{|u|}-\frac{3\, n_2}{|u|\beta_1}$  depend essentially on the lattice parameter $\beta_1$.
Here $n_{2,1}=N_{2,1}/L$ are  the dimensionless  densities  of pairs and  unpaired fermions, with the sound velocities $v_{1,2}$ given in (\ref{velocity}).
The study of the FFLO pair correlation is presented elsewhere \cite{SC}.
To conclude, we note that our work provides benchmark physics of the 1D attractive Hubbard model of relevance to experiments with ultracold  fermionic atoms on  lattices.

{\em Acknowledgments.}  The authors SC and YCY contributed equally to the calculations in this paper.
The authors  thank R. Hulet  for helpful discussion. This work is supported by
Key NNSFC grant number 11534014, MOST grant number 2017YFA0304500,
NNSFC grant numbers 11374331, 11174375 and ARC Discovery Projects DP130102839, DP170104934.

\setcounter{figure}{0}
\setcounter{equation}{0}
\def\thefigure{A\arabic{figure}}
\def\thetable{A\arabic{table}}
\def\theequation{A\arabic{equation}}
\def\thelabel{A\arabic{label}}
%%\setcounter{page}{1}
%\pagestyle{plain}

%%%%%%%%%%%%%%%%%%%%%%%%%%%%%%%%%%%%%%%%%%%%%%%%%%%%%%
%\appendix
%\appendix{\bf Appendix}
%%%%%%%%%%%%%%%%%%%%%%%%%%%%%%%%%%%%%%%%%%%%%%%%%%%%%%

\section{Appendices}

\subsection{Wiener-Hopf method}
\label{app-wienerhopf}

The phase boundary between phases IV and V is determined by the conditions $\varepsilon^u(0)=0$ and $\varepsilon^b(0)< 0$,
which imply that $Q=0$ and $A$ is finite.
Thus at zero temperature the TBA equations are simplified to
\begin{align}
\varepsilon^u(k)=&-2\cos k-\mu-2u-B \notag\\
&-\int_{-A}^A \textmd{d}\Lambda \, a_1(\sin k-\Lambda){\varepsilon_1^\prime}(\Lambda), \label{ma-ra14}\\
\varepsilon_1^\prime(\Lambda)=&-2\mu-2\int_{-\pi}^\pi \textmd{d}k \, \cos^2 k  \, a_1(\sin k -\Lambda) \notag\\
&-\int_{-A}^A \textmd{d}\Lambda^\prime \, a_2(\Lambda-\Lambda^\prime) {\varepsilon_1^\prime}(\Lambda^\prime). \label{ma-ra15}
\end{align}
Particularly, if $A=\infty$, it follows that the chemical potential  $\mu=0$.
The intersection of the phase boundary with  the $B$-axis  could be calculated exactly by the  Fourier transformation
\begin{align}
B_{c1}=2\left|u\right|-2+2\int_0^\infty \textmd{d}\omega \, \frac{J_1(\omega)\exp(-|u|\omega)}{w\cosh(u\,\omega)}.
\end{align}

Now we  consider the more general case $A \gg 1$, for  which  the phase boundary can be resolved using the Wiener-Hopf method.
By applying Fourier transformation on \cref{ma-ra15} and after some algebraic manipulations, we have
\begin{widetext}
\begin{equation}
\varepsilon_1^\prime(\Lambda) = -\mu-\int_{-\infty}^\infty \textmd{d}\omega \, \frac{J_1(\omega)}{\omega \cosh(u\, \omega)} \exp(i\,\omega\,\Lambda)
+\int_0^\infty \textmd{d}\Lambda^\prime \, \varepsilon_1^\prime(\Lambda^\prime+A) \, \left[R(\Lambda-\Lambda^\prime-A)+R(\Lambda+\Lambda^\prime+A)\right], \label{ma-ra1}
\end{equation}
\end{widetext}
where we have introduced the function
\begin{align}
R(x)=\int_{-\infty}^\infty \frac{\textmd{d}w}{2\pi} \,\frac{\exp(\mathrm{i} \,w\,x)}{1+\exp(2|u\,w|)}.
\end{align}

Substituting  $y(\Lambda)=\varepsilon_1^\prime(\Lambda+A)$ and expanding  $y(\Lambda)=\sum_{n=0}^\infty y_n(\Lambda)$
in terms of powers of $\Lambda$ in \cref{ma-ra1}, the result can be  separated
 into a series of Wiener-Hopf integral equations in terms of the functions  $y_n(\Lambda)$, namely
\begin{align}
y_n(\Lambda)=g_n(\Lambda)+\int_0^\infty \textmd{d}\Lambda^\prime \, R(\Lambda-\Lambda^\prime) \, y_n(\Lambda^\prime). \label{ma-yn}
\end{align}
Here we denote the driving terms
\begin{eqnarray}
g_0(\Lambda)&=&-\mu-\int_{-\infty}^\infty \textmd{d}\omega \, \frac{J_1\left(\omega\right) \mathrm{e}^{\mathrm{i}\omega\left(\Lambda+A\right)}}{\omega \cosh\left(u\, \omega\right)}, \notag\\
g_n(\Lambda)&=& \int_0^\infty \textmd{d}\Lambda^\prime \, R(\Lambda+\Lambda^\prime+2A) \, y_{n-1}(\Lambda^\prime).
\end{eqnarray}

To solve these integral equations for $y_n(\Lambda)$, we begin by defining
\begin{align}
\tilde{y}_n^\pm(\omega)=&\int_{-\infty}^\infty \textmd{d}\Lambda \,\theta_H\left(\pm \Lambda\right)\, y_n\left(\Lambda\right) \mathrm{e}^{\mathrm{i}\omega\Lambda}, \notag
\end{align}
where $\tilde{y}_n^+(\omega)$ ($\tilde{y}_n(\omega)$) is an analytic function in  the upper (lower) half-plane.
It is  obvious that  the Fourier transformation of $y_n(x)$ satisfies  the relation $\tilde{y}_n(\omega)=\tilde{y}_n^+(\omega)+\tilde{y}_n^-(\omega)$.

From \cref{ma-yn} it follows that
\begin{align}
\tilde{y}_n^+(\omega) \, \frac{1}{1+\exp(-2|u|\,|\omega|)} + \tilde{y}_n^-(\omega)=\tilde{g}_n(\omega),\label{ma-ra17}
\end{align}
by applying Fourier transformation. We further decompose the denominator  $1+\exp(-2|u|\,|\omega|)$ into a product of two pieces,
\begin{align}
1+\exp(-2|u|\, |\omega|)=G^+(\omega)G^-(\omega),
\end{align}
where $G^+(\omega)$ ($G^-(\omega)$) is an analytic function in the upper (lower) half-plane.
Then substituting this last equation into \cref{ma-ra17} results in the  form
\begin{align}
\tilde{y}_n^+(\omega)/G^+(\omega)+G^-(\omega)\tilde{y}_n^-(\omega)=G^-(\omega)\tilde{g}_n(\omega).\label{ma-ra18}
\end{align}

Furthermore, we decompose  $G^-(\omega)\tilde{g}_n(\omega)$ into a sum of two pieces,
\begin{align}
G^-(\omega)\tilde{g}_n(\omega)=Q_n^+(\omega)+Q_n^-(\omega), \label{product}
\end{align}
where similarly $Q_n^+(\omega)$ ($Q_n^-(\omega)$) is an analytic function in the upper (lower) half-plane.
Then substitution of this last equation into \cref{ma-ra18} gives
\begin{align}
\tilde{y}_n^+(\omega)=&\,G^+(\omega)Q_n^+(\omega),\label{ma-ra18.1}\\
\tilde{y}_n^-(\omega)=&\,Q_n^-(\omega)/G^-(\omega).
\end{align}
In this way we can  work out the Fourier transformation of $y_0(\Lambda)$ and  $y_n(\Lambda)$ itself.

To this end, recalling (\ref{product}), we firstly decompose $1+\exp(-2|u|\, |\omega|)$ as
\begin{align}
&G^+(\omega)=G^-(-\omega) \notag\\
=&\frac{\sqrt{2\pi}}{\Gamma(\frac{1}{2}-  \frac{\mathrm{i}  |u|  \omega}{\pi})}
\left( -\frac{\mathrm{i}  |u|\omega}{\pi} \right)^{-\frac{\mathrm{i} |u|\omega}{\pi}}\exp\left(\frac{\mathrm{i} |u|\omega}{\pi}\right),
\end{align}
where we should note that $\lim_{\omega \rightarrow \infty} G^{\pm}(\omega)=1$, along with the special values
$G^\pm(0)=\sqrt{2}$ and $G^\pm\left(\pm \frac{\mathrm{i} \pi}{2|u|}\right)=\sqrt{{\pi}/{\mathrm{e}}}$ of these functions.

The decomposition for $G^-(\omega)\tilde{g}_n(\omega)$ in general is subtle,
however the leading case $G^-(\omega)\tilde{g}_0(\omega)$ is accessible.
We start analysis from the Fourier transformation of $g_0(\Lambda)$,
\begin{align}
\tilde{g}_0(\omega)=-\mu\, 2\pi \delta_D(\omega)-\frac{2\pi J_1(\omega)\exp(-\mathrm{i} \, \omega A)}{\omega \cosh(u\omega)}, \label{ma-ra19}
\end{align}
where on the rhs the $\delta_D$ function could be decomposed as
\begin{align}
2\pi \delta_D(\omega)=\mathrm{i} \left( \frac{1}{\omega+\mathrm{i} \, \epsilon}-\frac{1}{\omega-\mathrm{i} \,\epsilon} \right) \qquad \left(\epsilon \rightarrow +0 \right).
\end{align}
The second term on the rhs is a meromorphic function of $\omega$ with poles located at
\begin{align}
\omega_n=\mathrm{i} \frac{\pi}{2|u|}(2n+1) \qquad \left(n \in \mathbf{Z}\right)
\end{align}
originating from the term $\frac{1}{\cosh(u\omega)}$, implying the decomposition
\begin{align}
\frac{1}{\cosh(u\omega)}=& \, \chi^+(\omega)+\chi^-(\omega), %\label{ma-ra20}
\notag\\
\chi^+(\omega)=&\frac{\mathrm{i}}{|u|}\sum_{n=0}^\infty (-1)^n \frac{1}{\omega+\omega_n}, %\label{ma-ra21}
\notag\\
\chi^-(\omega)=&\frac{1}{\cosh(u\omega)}-\frac{\mathrm{i}}{|u|}\sum_{n=0}^\infty (-1)^n \frac{1}{\omega+\omega_n}, \label{ma-ra22}
\end{align}
where $\chi^+(\omega)$ and $\chi^-(\omega)$ are analytic functions in the upper and lower half-planes, respectively.
With the help of \cref{%ma-ra20,ma-ra21,
ma-ra22}, as for any analytic and bounded function $f^-(\omega)$ in the lower half-plane, the decomposition of $\frac{f^-(\omega)}{\cosh(u \omega)}$ is
\begin{align}
\frac{f^-(\omega)}{\cosh(u \omega)}=&\,F^+(\omega)+F^-(\omega),%\label{decom-1}
\notag\\
F^+(\omega)=&\,\frac{\mathrm{i}}{|u|}\sum_{n=0}^\infty (-1)^n \frac{f^-(-\omega_n)}{\omega+\omega_n},%\label{decom-2}
\notag\\
F^-(\omega)=&\,\frac{f^-(\omega)}{\cosh(u \omega)}-F^+(\omega)\label{decom-3}.
\end{align}

By virtue of \cref{ma-ra19,%decom-1,decom-2,
decom-3}, we make the following decomposition for $G^-(\omega)\tilde{g}_0(\omega)$,
\begin{align}
Q_0^+(\omega)=&-\frac{\mathrm{i} \, \mu\, G^-(0)}{\omega+\mathrm{i} \,\epsilon}-q(\omega)\\
Q_0^-(\omega)=&\frac{\mathrm{i} \, \mu\, G^-(0)}{\omega+\mathrm{i} \,\epsilon}
- \frac{2\pi J_1(\omega)\exp(-\mathrm{i} \,\omega A)G^-(\omega)}{\omega \cosh(u \omega)} + q(\omega) \notag
\end{align}
where $q(\omega)=4\mathrm{i} \sum_{n=1}^\infty (-1)^n \frac{G^-(-\mathrm{i} h_n)I_1(h_n)\exp(-h_n A)}{(2n+1)(\omega+\mathrm{i} h_n)}$, $I_1(z)$
is the first order modified Bessel function, $h_n=\frac{\pi}{2|u|}(2n+1)$, with the series converging only if $A>1$.

If $A \gg 1$, using \cref{ma-ra18.1}, we have
\begin{align}
y_0^+(\omega)=G^+(\omega)\left[ -\frac{\mathrm{i} \, \mu G^-(0)}{\omega+\mathrm{i} \, \epsilon}- q(\omega) \right]. \label{ma-ra23}
\end{align}
Obviously, we know $y(0)=\varepsilon_1^\prime(A)=0$, which implies
\begin{align}
0=y(0)=\lim_{\omega \rightarrow \infty} -\mathrm{i} \,\omega \, \tilde{y}^+(\omega). \label{ma-ra24}
\end{align}
Hereafter we replace $y(\omega)$ with $y_0(\omega)$, which is a reasonable approximation if $A \gg 1$. Therefore \cref{ma-ra23,ma-ra24} give rise to
\begin{align}
\mu=-4\sum_{n=0}^\infty \frac{G^-(-\mathrm{i} \,h_n)I_1(h_n)\exp(-h_n A)}{(2n+1)G^-(0)}. \label{ma-pb45-chem}
\end{align}

Since  we have obtained a parametric expression for the critical chemical potential, we turn  to the expression for the magnetic field.
Due to the fact that  the phase boundary is determined by $\varepsilon^u (0)=0$, we thus use  \cref{ma-ra14} to determine the magnetic field.

For simplicity, we rewrite \cref{ma-ra14,ma-ra15} as
\begin{align}
\varepsilon^u(k)=&-2\cos k-\mu-2u-B \notag\\
&+ \int_A^\infty \textmd{d}\Lambda \, \left[ a_1(\sin k - \Lambda) + a_1(\sin k + \Lambda) \right] \varepsilon^{\prime}_1(\Lambda) \notag \\
&-\int_{-\infty}^\infty \textmd{d}\Lambda \, a_1(\sin k - \Lambda) \varepsilon_1^\prime(\Lambda), \label{re-tba-1}\\
\varepsilon_1^\prime(\Lambda)=&\,\varepsilon_1^{\prime (0)}(\Lambda) - \int_{-A}^A \textmd{d}\Lambda^\prime \, a_2(\Lambda-\Lambda^\prime) \varepsilon_1^\prime(\Lambda^\prime), \label{re-tba-2}
\end{align}
where we have denoted
\begin{align}
\varepsilon_1^{\prime(0)}(\Lambda)=-2\mu-2\int_{-\pi}^\pi \textmd{d}k \, \cos^2 k  \, a_1(\sin k - \Lambda). \label{ma-ra25}
\end{align}

Substituting  \cref{re-tba-2} into the last term on the rhs of \cref{re-tba-1} gives
\begin{align}
&\varepsilon^u(k) \notag\\
=&-2\cos k-\mu-2u-B \notag\\
&+\int_0^\infty \textmd{d}\Lambda \, \left[s(\Lambda+A-\sin k)+ s(\Lambda+A+\sin k) \right] \, y(\Lambda) \notag\\
&-\int_{-\infty}^\infty \textmd{d}\Lambda \, s(\Lambda-\sin k) \varepsilon_1^{\prime (0)}(\Lambda)\label{ma-ra29},
\end{align}
where we have introduced the function $s(x)=\frac{1}{4|u|\cosh(\frac{\pi x}{2|u|})}$ and made use of the two identities
\begin{eqnarray}
\frac{1}{4|u|\cosh(\frac{\pi x}{2|u|})}&=&\sum_{n=0}^\infty (-1)^n a_{2n+1}(x), \notag\\
\int_{-\infty}^\infty \textmd{d}y \, a_n(x-y) a_m(y-z)&=&\,a_{m+n}(x-z).\notag
\end{eqnarray}

Substituting the expansion $s(x)=\frac{1}{2|u|}\sum_{n=0}^\infty (-1)^n \exp(-h_n x)$,
where $\left| {\pi x}/{u} \right| < 1$ and \cref{ma-ra25} into \cref{ma-ra29},
and after some algebraic manipulations, we arrive at the result
\begin{align}
\varepsilon^u(k)=&-2\cos k-2u-B \label{ma-ra32}\\
&+\sum_{n=0}^\infty \frac{(-1)^n}{|u|} \, \tilde{y}^+(\mathrm{i} \,h_n) \cosh(h_n \sin k) \exp(-h_n A) \notag\\
&+ 2\int_0^\infty \frac{  \textmd{d}\omega J_1(\omega) \cos(\omega \sin k) \exp(-|u|\omega) }{\omega \cosh(u\omega)}, \nonumber
\end{align}

Using \cref{ma-ra32} and $\varepsilon^u (0)=0$ we derive the  expression
\begin{align}
B=&-2+2|u|+\sum_{n=0}^\infty \frac{(-1)^n}{|u|} \, \tilde{y}^+(\mathrm{i} \,h_n) \exp(-h_n A) \notag\\
&+2\int_0^\infty \textmd{d}\omega \frac{ J_1(\omega) \exp(-|u|\omega) }{\omega \cosh(u\omega)} \label{ma-pb45-mag}
\end{align}
for determining the critical magnetic field.
Here we denoted  $h_n= \frac{\pi}{2|u|}(2n+1)$.
The equation (\ref{ma-pb45-mag}) sets up a relation between the magnetic field and the chemical potential.
In summary, the phase boundary between phase IV and V is determined  by \cref{ma-pb45-chem,ma-pb45-mag}  for  $A \gg 1$.

\def\theequation{B\arabic{equation}}

\subsection{ Derivation of the Equation of State }
\label{app-eos}

The derivation of the equation of state is rather involved.
Here we sketch the calculations for the terms $p^u$ and $p^b_n$.

Prior to substituting the dressed energies into the definitions of $p^u$ and $p^b_n$ integrating by parts,
we first need to find a suitable form of the TBA equations for this procedure, i.e., (\ref{reduced-tba-u}) and (\ref{reduced-tba-b}).
For simplicity in later discussion, we approximate the definition of $p^b_n$ as
\begin{align}
p^b_n=&\,\int_{-\infty}^\infty \frac{\textmd{d}\Lambda}{\pi} \textmd{Re} \frac{1}{\sqrt{\Lambda+\mathrm{i} \, n \, |u|}} \varepsilon_n^{\prime -}(\Lambda)  \notag\\
=&\,T\int_{-\infty}^\infty \frac{\textmd{d}\Lambda}{2\pi} \int_{-\pi}^\pi \textmd{d}k \, a_n(\Lambda-\sin k) \notag\\
=&\,\int_{-\infty}^\infty \textmd{d}\Lambda \,\varepsilon_n^{\prime -}(\Lambda)  \Delta_n(\Lambda) + o\left( \frac{1}{|u|^4} \right), \label{p-b-n-expansion}
\end{align}
where $\Delta_n(\Lambda)=a_n(\Lambda)-\frac{1}{2} b_n(\Lambda)+ 2\Lambda^2 b_n(\Lambda)$.
In the above equations, we used the abbreviations
\begin{eqnarray}
\varepsilon_n^{\prime -}(x)& =&T\ln\left(1+\mathrm{e}^{-\varepsilon_n^\prime(x)/T}\right),\nonumber\\
\varepsilon_n^{ -}(x) &=&T\ln\left(1+\mathrm{e}^{-\varepsilon_n(x)/T}\right).\nonumber
\end{eqnarray}

To obtain the result (\ref{reduced-tba-u}), regarding the first series of integral terms on the rhs of (\ref{tba1}),
we expand them in the strong coupling regime as
\begin{align}
&\sum_{n=1}^\infty  \int_{-\infty}^\infty \textmd{d}\Lambda \, a_n(\sin k-\Lambda)\varepsilon_n^{\prime -}(\Lambda) \notag\\
=&\sum_{n=1}^\infty  \int_{-\infty}^\infty \textmd{d}\Lambda \, \Delta_n(\Lambda) \varepsilon_n^{\prime -}(\Lambda)  \times \notag\\ &\frac{a_n(\Lambda)\left\{ 1+ \frac{2\Lambda \sin k - \sin^2 k}{(nu)^2+\Lambda^2} + \left[\frac{2\Lambda \sin k - \sin^2 k}{(nu)^2+\Lambda^2} \right]^2 \right\}}{\Delta_n(\Lambda)} + o\left( \frac{1}{|u|^4} \right) \notag\\
=& \sum_{n=1}^\infty p_n^b + \bar{a} + 2\bar{a} \cos^2 k + o\left( \frac{1}{|u|^4} \right),
\end{align}
where we have inserted (\ref{p-b-n-expansion}) and $\bar{a}$ has been defined in \cref{sec-eos}.

While for the second series of integral terms, it is easy to see that under the assumption $B/T \gg 1$,
these spin-wave contributions are no more than $-T \,\mathrm{e}^{-2B/T} \mathrm{e}^{-\bar{K}} I_0(\bar{K})$,
which is accessible through simple iteration of (\ref{tba2}).
In fact, the spin degree of freedom is frozen here, and thus this term could be neglected in later discussion.

We can rewrite (\ref{reduced-tba-u}) as
\begin{align}
\varepsilon^u(k)=\varepsilon^u_0(k)-A^u,
\end{align}
where $\varepsilon^u_0(k)=-2\cos k + 2 \bar{a} \cos^2 k$ and $A^u=\mu+2u+B-\sum_{n=1}^\infty p^b_n + \bar{a}$.

Integrating by parts in $p^u$, we obtain
\begin{align}
p^u=&T \ln\left( 1+\mathrm{e}^{(\mu+2u+B-\sum_{n=1}^\infty p_n^b + \bar{a} -2)/T} \right)\notag\\
&+\frac{1}{\pi}\int_{2\bar{a}-2}^{2\bar{a}+2} \frac{\textmd{d}\varepsilon^u_0\,k(\varepsilon_0^u)}{1+\mathrm{e}^{\varepsilon_0^u/T}/z},
\end{align}
where $z=\mathrm{e}^{A^u/T}$ and $k(\varepsilon_0^u)=\arccos \left( \frac{1-\sqrt{1+2\bar{a}\varepsilon_0^u}}{2\bar{a}} \right)$ represents the inverse function of $\varepsilon_0^u (k)$.
By taking account of $\bar{a} \sim \sum_{n=1}^\infty \frac{p^b_n}{(nu)^2}$ for the  strong coupling regime, the integral in the above equation can be further simplified,
\begin{eqnarray}
&& \frac{1}{\pi}\int_{2\bar{a}-2}^{2\bar{a}+2} \frac{\textmd{d}\varepsilon^u_0\,k(\varepsilon_0^u)}{1+\mathrm{e}^{\varepsilon_0^u/T}/z}
=\frac{2}{\pi}\int_{\bar{a}-1}^{\bar{a}+1} \textmd{d}x\, \frac{k(2x)}{1+\mathrm{e}^{2x/T}/z} \notag\\
&&=  \,\frac{2}{\pi}\int_{-1}^{1}\textmd{d}x\, \frac{\arccos(-x)}{1+\mathrm{e}^{2x/T}/z}-\frac{2\bar{a}}{\pi}\int_{-1}^{1} \frac{\textmd{d}x\, x^2/\sqrt{1-x^2}}{1+\mathrm{e}^{2x/T}/z} \notag\\
&& \quad +\frac{2\bar{a}}{1+\mathrm{e}^{2\varepsilon^u(\pi)/T}}+o\left(\frac{1}{|u|^4}\right),
\end{eqnarray}
where we have changed the integration variable $\varepsilon^u_0=2x$
and then applied Taylor expansion with respect to $\bar{a}$.
See $\varepsilon^u(\pi)$ in \cref{sec-eos}. The result (\ref{pressure-a1}) is therefor achieved.

We then turn to the transformation of $\varepsilon_n^\prime(\Lambda)$. Similar to the treatment for $\varepsilon^u(k)$,
we employ Taylor expansion to expand (\ref{tba3}) in  the strong coupling region, with result
\begin{eqnarray}
\varepsilon^\prime_n(\Lambda)&=&-2n\mu- a_n(\Lambda) \left[ 2\pi-\int_{-\pi}^\pi \textmd{d}k\, \cos k \varepsilon^{u -}(k)   \right] \notag\\
&& - b_n(\Lambda) \left[ -\frac{\pi}{2}-\int_{-\pi}^\pi \textmd{d}k\, \cos k \, \sin^2 k \,\varepsilon^{u -}(k) \right] \notag\\
&&+ \sum_{m=1}^\infty T_{nm}\ast \varepsilon_m^{\prime -}(\Lambda)   + o\left( \frac{1}{|u|^4} \right),
\end{eqnarray}
where the integral terms in the brackets are denoted as $d_1$ and $d_2$, respectively.
Here $d_1$ and $d_2$ can be calculated via integration by parts, similar to that done for $p^u$ above,
see the explicit expressions in \cref{sec-eos}.
With respect to the convolution term, due to the condition of low density,
the cut-off of the dressed energy $\varepsilon^\prime_n(\Lambda)$ is small, thus in general we can make the approximations
\begin{eqnarray}
& &\int_{-\infty}^\infty \textmd{d}\Lambda^\prime \, a_p(\Lambda-\Lambda^\prime)\varepsilon_q^{\prime -}(\Lambda^\prime)  \notag\\
&=&\int_{-\infty}^\infty \textmd{d}\Lambda^\prime \, a_p(\Lambda^\prime) \varepsilon_q^{\prime -}(\Lambda^\prime) \notag\\
&&-\Lambda^2 \int_{-\infty}^\infty \textmd{d}\Lambda^\prime \, b_p(\Lambda^\prime) \varepsilon_q^{\prime -}(\Lambda^\prime) +o \left( \frac{1}{|u|^4} \right),
\end{eqnarray}
which results in \cref{reduced-tba-b} in the main text.

We next rewrite (\ref{reduced-tba-b}) as
\begin{align}
\varepsilon^\prime_n(\Lambda)=D_n \left(\frac{\Lambda}{n|u|}\right)^2 - A^b_n,
\end{align}
where $A^b_n=2n\mu-\eta_n+\frac{d_1}{\pi n |u|}+\frac{d_2}{2\pi(n|u|)^3}$ and $D_n$ was defined in \cref{sec-eos}.
\cref{pressure-a2} is arrived at by substituting the above equation into the definition of $p^b_n$ and integrating by parts.

Lastly, $\bar{a}$,
$\xi_p^m=T\int_{-\infty}^\infty \textmd{d}\Lambda^\prime \, a_p(\Lambda^\prime) \varepsilon_m^{\prime -}(\Lambda^\prime)$
and $\phi_p^m=T\int_{-\infty}^\infty \textmd{d}\Lambda^\prime \, b_p(\Lambda^\prime) \varepsilon_m^{\prime -}(\Lambda^\prime)$ are calculated in a similar way.

\def\theequation{C\arabic{equation}}

\subsection{Some constants in the scaling functions}
\label{Scaling-Constants}

The constants used  in the  scaling forms for the phase transition (II-IV)  are given explicitly by
\begin{eqnarray}
\lambda_1&=&-\frac{2\sqrt{|u|}(1-q_{c4}/\pi)}{\sqrt{2\pi-q_{c4}^3/3}}, \notag\\
\lambda_2&=&\frac{(1-q_{c4}/\pi)q_{c4}/\pi}{\sqrt{|u|}\sqrt{2\pi-q_{c4}^3/3}},\notag\\
n_{b4}&=& \gamma, \quad m_{b4}=\frac{1}{2}\gamma,\nonumber\\
\kappa_{b4}&=&\gamma' \left(1+ \frac{4}{\sqrt{\pi}} \tau^{\frac{1}{2}} \tilde{f}_{\frac{1}{2}} + \frac{6}{\pi} \tau \tilde{f}_{\frac{1}{2}}^2 + \frac{5}{\pi^{\frac{3}{2}}}\tau^{\frac{3}{2}} \tilde{f}_{\frac{1}{2}}^3 - \frac{2}{\sqrt{\pi}} \tau^{\frac{3}{2}} \tilde{f}_{\frac{3}{2}}  \right), \notag\\
\lambda_3&=&-\frac{4\sqrt{|u|}(1-q_{c4}/\pi)}{\sqrt{2\pi-q_{c4}^3/3}}, \notag\\
\chi_{b4}&=&\frac{1}{2} \gamma^{\,\prime} + \frac{2\delta \, \gamma^{\,\prime}-(1-\gamma)\delta'}{|u|\sqrt{\pi}}\left( \tau^{\frac{1}{2}} \tilde{f}_{\frac{1}{2}} + \frac{1}{2\sqrt{\pi}} \tau \tilde{f}_{\frac{1}{2}}^2 + \frac{1}{4\pi}\tau^{\frac{3}{2}} \tilde{f}_{\frac{1}{2}}^3 \right. \notag\\
&&\left.- \tau^{\frac{3}{2}} \tilde{f}_{\frac{3}{2}} \right), \notag\\
\lambda_4&=&-\frac{2(q_{c4}/\pi)^2(1-q_{c4}/\pi)}{\sqrt{|u|^3}\sqrt{2\pi-q_{c4}^3/3}}.
\end{eqnarray}
Here the parameter $q_{c4}=\sqrt{B+2-2\sqrt{1+u^2}} +\frac{1}{3\pi|u|}\left( B+2-2\sqrt{1+u^2}\right)$.

The constants used  in the  scaling forms for the phase transition (V-IV)  are given explicitly by
\begin{eqnarray}
n_{b5}&=&-\frac{2}{\sqrt{\pi}}\tau^{\frac{1}{2}} \tilde{f}_{\frac{1}{2}} -\frac{1}{\pi}\tau \tilde{f}_{\frac{1}{2}}^2 - \frac{1}{2\pi^{\frac{3}{2}}}\tau^{\frac{3}{2}} \tilde{f}_{\frac{1}{2}}^3 + \frac{1}{\sqrt{\pi}}\tau^{\frac{3}{2}} \tilde{f}_{\frac{3}{2}}, \notag\\
\lambda_5&=& -\frac{1}{2\sqrt{\pi}} \left( 1- \frac{4}{\pi}\sqrt{1+\frac{2\pi|u|\tilde{\mu}_{c5}}{2\pi-q_{c5}^3/3}}\right),\notag\\
\lambda_6&=&-\frac{1}{2\sqrt{\pi}} \left( 1 - \frac{8}{\pi} \sqrt{1+\frac{2\pi|u|\tilde{\mu}_{c5}}{2\pi-q_{c5}^3/3}} \right)\nonumber\\
\kappa_{b5}&=& -\frac{\left(1-\gamma\right) \tilde{f}_{-\frac{1}{2}}}{D_0 \tau^{\frac{1}{2}}}  \left( \frac{4}{\sqrt{\pi}} + \frac{6}{\pi}\tau^{\frac{1}{2}} \tilde{f}_{\frac{1}{2}} + \frac{6}{\pi^{\frac{3}{2}}}\tau \tilde{f}_{\frac{1}{2}}^2  \right.\notag\\
&&\left.- \frac{7}{4\pi}\tau^{\frac{3}{2}} \tilde{f}_{\frac{3}{2}}+ \frac{5}{\pi^2} \tau^{\frac{3}{2}} \tilde{f}_{\frac{1}{2}}^3 \right),\\ \notag
\end{eqnarray}
where  $\tilde{\mu}_{c5} \approx 2|u|-B-2+\frac{8\sqrt{2}}{3\pi|u|\alpha_1}\left(2\sqrt{1+u^2}-B-2\right)^{\frac{3}{2}}$
and $q_{c5}=\sqrt{\tilde{\mu}_{c5}+2u+B+2}$.

\def\theequation{D\arabic{equation}}
\subsection{ Explicit forms of the additivity rules }
\label{app-iteration}

The vectorial forms of \cref{vec-tba1,vec-tba2} are accessible by expanding the rescaled TBA equations in terms of
$\left\{k^{2n}\right\}$ and $\left\{\Lambda^{2n}\right\}$ ($n=0,1,2\ldots$).
We here give the explicit expression for the matrix $\mathbf{A}^n(\tilde{y}_c)$ ($n=1,2$) with the elements
\begin{align}
\left\{ \mathbf{A}^n (\tilde{y}_c) \right\}_{jl}=\frac{2}{\pi} \sum_{0\leq j \leq i < \infty} \frac{(-1)^i \, C_{2i}^{2j} \, \tilde{y}_c^{2i-2j+2l+1}}{n^{2i+1}(2i-2j+2l+1)},
\end{align}
with $j,l=0,1,2,\ldots$.
Thus the first two orders of $\mathbf{A}_{(q)}^n (y_c)$ are written as
\begin{widetext}
\begin{align}
\mathbf{A}^1_{(1)}(\tilde{y}_c)=&\frac{2}{\pi}
\left[
\begin{array}{cccc}
 1 & 0  & \cdots \\
 -1& 0 & \cdots \\
 \vdots &  &  &  \\
\end{array}
\right]
\, \tilde{y}_c,
&
\mathbf{A}^2_{(1)}(\tilde{y}_c)=&\,\frac{1}{\pi}
\left[
\begin{array}{cccc}
 1 & 0 & \cdots \\
 -\frac{1}{4} & 0 & \cdots \\
 \vdots &  &  &  \\
\end{array}
\right]
\, \tilde{y}_c,
\label{an_1_matrix} \\
\mathbf{A}^{1}_{(3)}(\tilde{y}_c)=&\frac{2}{\pi}
\left[
\begin{array}{cccc}
 -\frac{1}{3} & \frac{1}{3}  & \cdots \\
 2 & \frac{1}{3}  & \cdots \\
 \vdots &  &  &  \\
\end{array}
\right]
\,\tilde{y}_c^3,
&
\mathbf{A}^2_{(3)}(\tilde{y}_c)=&\frac{1}{\pi}
\left[
\begin{array}{cccc}
 -\frac{1}{12} & \frac{1}{3} & \cdots \\
  \frac{1}{8}  & -\frac{1}{12} & \cdots \\
 \vdots &  &  &  \\
\end{array}
\right]
\, \tilde{y}_c^3.
\label{an_3_matrix}
\end{align}
\end{widetext}

Next, the partial derivatives of \cref{vec-tba1,vec-tba2} read
\begin{align}
\frac{\partial \vec{\varepsilon}^{\,1}}{\partial \tilde{\mu}}
=&\frac{\partial \vec{V}^1}{\partial \tilde{\mu}}
-\mathbf{A}^1(\tilde{\Lambda}_c)\,\frac{\partial \vec{\varepsilon}^{\,2}}{\partial \tilde{\mu}},\label{vec-tba1-dmu}\\
\frac{\partial \vec{\varepsilon}^{\,2}}{\partial \tilde{\mu}}
=&\vec{V}^2
-\mathbf{A}^1(\tilde{k}_c)\,\frac{\partial \vec{\varepsilon}^{\,1}}{\partial \tilde{\mu}}
-\mathbf{A}^2(\tilde{\Lambda}_c)\,\frac{\partial \vec{\varepsilon}^{\,2}}{\partial \tilde{\mu}}.\label{vec-tba2-dmu}
\end{align}
With the help of the explicit forms of  $\mathbf{A}^n (\tilde{y}_c)$, i.e., (\ref{an_1_matrix}) and (\ref{an_3_matrix}), we can
obtain \cref{density-by-Fermi-point}, which relates the densities and the cutoffs.
Together with \cref{solution_mu_matrix_hubbard}, we then obtain the relation between the densities and the effective chemical potentials (\ref{chemical-1}) and (\ref{chemical-2}).

On this basis, we now proceed to  derive the explicit expressions for the effective compressibility and susceptibility
in terms of densities of bound pairs and excess fermions.
Apparently, the densities of bound pairs and excess fermions rely on the chemical potential and the magnetic field, and vice versa,
which in fact indicates under fixed magnetic field one could obtain the following results through the total derivatives,
\begin{align}
\kappa_1=\left( \frac{\partial n_1}{\partial \mu_1} \right)_B=\frac{\textmd{d}n_1}{\textmd{d}\mu}, \,\,
\kappa_2=2\left( \frac{\partial n_2}{\partial \mu_2} \right)_B=\alpha_1\frac{\textmd{d}n_2}{\textmd{d}\mu},
\end{align}
where we keep $d\,B=\frac{\partial B}{\partial n_1}d\,n_1+\frac{\partial B}{\partial n_2}d\,n_2=0$. Thus we have
\begin{align}
\frac{\textmd{d}n_1}{\textmd{d}\mu}=\frac{1}{J} \left( \frac{\partial B}{\partial n_2} \right)_{n_1}, \quad
\frac{\textmd{d}n_2}{\textmd{d}\mu}=-\frac{1}{J} \left( \frac{\partial B}{\partial n_1} \right)_{n_2}.
\end{align}
Here the Jacobian determinant
\begin{align}
J=&\left( \frac{\partial \mu}{\partial n_1} \right)_{n_2} \left( \frac{\partial B}{\partial n_2} \right)_{n_1} - \left( \frac{\partial B}{\partial n_1} \right)_{n_2} \left( \frac{\partial \mu}{\partial n_2} \right)_{n_1} \notag\\
=&-\frac{\alpha_1}{2} \left[ \left( \frac{\partial \mu_1}{\partial n_1} \right)_{n_2} \left( \frac{\partial \mu_2}{\partial n_2} \right)_{n_1} - \left( \frac{\partial \mu_2}{\partial n_1} \right)_{n_2} \left( \frac{\partial \mu_1}{\partial n_2} \right)_{n_1} \right],
\end{align}
where we have used \cref{effective-mu1,effective-mu2}.

Similarly, the magnetic field is dependent on the effective chemical potentials while the latter is dependent on densities of bound pairs and excess fermions.
Therefore by application of chain rule we have
\begin{align}
\left(\frac{\partial B}{\partial n_1}\right)_{n_2}=&\left(\frac{\partial \mu_1}{\partial n_1}\right)_{n_2}
-\frac{\alpha_1}{2}\left(\frac{\partial \mu_2}{\partial n_1}\right)_{n_2}, \notag\\
\left(\frac{\partial B}{\partial n_2}\right)_{n_1}=&\left(\frac{\partial \mu_1}{\partial n_2}\right)_{n_1}
-\frac{\alpha_1}{2}\left(\frac{\partial \mu_2}{\partial n_2}\right)_{n_1}.
\label{dB-dn}
\end{align}
It is obvious that once the explicit expression of $\mu_r$ in terms of $n_s$ ($r,s=1,2$) is known,
our goal of the effective compressibilities is easy to achieve.
We use \cref{solution_mu_matrix_hubbard,definition_T_matrix,Fermi-point-dens} to derive
\begin{align}
\mu_1=&\, \pi^2 n_1^2\left[1+2\left(\frac{2n_2}{|u|}\right)+3\left(\frac{2n_2}{|u|}\right)^2\right] \notag\\ &+\frac{4\pi^2\alpha_1}{3\beta_1^3|u|}n_2^3\left[1+3\frac{2n_1+n_2}{\beta_1|u|}\right],\label{mu_in_n_up23_1} \\
\mu_2=&\, \frac{\pi^2 n_2^2}{\beta_1^2}\left[1+2\frac{2n_1+n_2}{\beta_1|u|}+3\left(\frac{2n_1+n_2}{\beta_1|u|}\right)^2\right] \notag\\
&+\frac{4\pi^2}{3\alpha_1 |u|}n_1^3\left[1+3\left(\frac{2n_2}{|u|}\right)\right] \notag\\
&+\frac{2\pi^2}{3\beta_1^3 |u|}n_2^3\left[1+3\frac{2n_1+n_2}{\beta_1|u|}\right],
\label{mu_in_n_up23_2}
\end{align}
and thus
\begin{align}
\kappa_1=&\frac{\pi^2}{J}\left[ -\frac{\alpha_1 n_2}{\beta_1^2}-\frac{4 \alpha_1 n_1 n_2}{|u|\beta_1^3}+\frac{4 n_1^2}{|u|}-\frac{4 n_1^3}{u^2}+\frac{24 n_1^2 n_2}{u^2} \right. \notag\\
&\left.-\frac{12 \alpha_1 n_1^2 n_2}{u^2 \beta_1^4}+\frac{6 \alpha_1 n_2^3}{u^2 \beta_1^4} \right], \notag \\
\kappa_2=&-\frac{2\alpha_1\pi^2}{J} \left[ n_1-\frac{n_1^2}{|u|}+\frac{4 n_1 n_2}{|u|}-\frac{6 n_1^2 n_2}{u^2}-\frac{\alpha_1 n_2^2}{u \beta_1^3} \right. \notag\\
&\left.+\frac{12 n_1 n_2^2}{u^2}-\frac{6 \alpha_1 n_1 n_2^2}{u^2 \beta_1^4} \right],\\
J=&-\frac{2\pi^4\alpha_1}{\beta_1^2}n_1 n_2
\left[
1+\frac{4 n_1}{|u|\beta_1}+\frac{12 n_1^2}{u^2 \beta_1^2}+\frac{4 n_2}{|u|}+\frac{4 n_2}{|u| \beta_1} \right. \notag\\
&\left.+\frac{24 n_1 n_2}{u^2 \beta_1^2}+\frac{8 n_1 n_2}{u^2 \beta_1}+\frac{12 n_2^2}{u^2}+\frac{10 n_2^2}{u^2 \beta_1^2}+\frac{16 n_2^2}{u^2 \beta_1}
\right].
\end{align}

The situation for effective susceptibilities is rather simple. With fixed total particle density, one confirms that $\textmd{d}n_1+2\textmd{d}n_2=0$,
and thus the total derivative of the effective chemical potentials with respect to $n_r$ ($r=1,2$) is
\begin{align}
\textmd{d} \mu_1
=&\left[ \left( \frac{\partial \mu_1}{\partial n_1}\right)_{n_2} -\frac{1}{2}\left( \frac{\partial \mu_1}{\partial n_2} \right)_{n_1} \right] \textmd{d} n_1, \notag\\
\textmd{d} \mu_2
=&-2\left[\left(\frac{\partial \mu_2}{\partial n_1}\right)_{n_2}
-\frac{1}{2}\left(\frac{\partial \mu_2}{\partial n_2}\right)_{n_1}\right] \textmd{d}n_2.
\end{align}
After some algebraic manipulations we then obtain
\begin{align}
\bar{\chi}_1=&1 \bigg/ \left(\frac{\partial \mu_1}{\partial n_1}
-\frac{1}{2}\frac{\partial \mu_1}{\partial n_2}\right), \notag \\
\bar{\chi}_2=&-1\bigg/ \left(\frac{\partial \mu_2}{\partial n_1}
-\frac{1}{2}\frac{\partial \mu_2}{\partial n_2}\right),
\end{align}
which together with \cref{mu_in_n_up23_1,mu_in_n_up23_2} result in
\begin{widetext}
\begin{align}
\bar{\chi}_1=&\frac{1/(2\pi^2)}{n_1-\frac{n_1^2}{|u|}+\frac{4 n_1 n_2}{|u|}-\frac{6 n_1^2 n_2}{u^2}-\frac{\alpha_1 n_2^2}{|u| \beta_1^3}+\frac{12 n_1 n_2^2}{u^2}-\frac{6 \alpha_1 n_1 n_2^2}{u^2 \beta_1^4}},\notag \\
\bar{\chi}_2=&\frac{1/\pi^2}{\frac{n_2}{\beta_1^2}-\frac{4 n_1^2}{|u| \alpha_1}+\frac{4 n_1^3}{u^2 \alpha_1 }+\frac{4 n_1 n_2}{|u| \beta_1^3}-\frac{24 n_1^2 n_2}{u^2 \alpha_1}+\frac{12 n_1^2 n_2}{u^2 \beta_1^4}-\frac{6 n_2^3}{u^2 \beta_1^4}}.
\end{align}
\end{widetext}


\begin{thebibliography}{99}
\bibitem{note1} The sound velocity here is accessible by using the formula $v_r=\sqrt{\frac{L}{m_p\, n_r \, r} \frac{\partial^2 E_r}{\partial L^2}}$,
where the particle mass has been rescaled as $m_p=\frac{1}{2}$ and $E_r=e_r\,L$. See $e_r$ in \cref{result_E1_hubbard,result_E2_hubbard}.
\end{thebibliography}
\end{document}